\documentclass[letterpaper]{article} % DO NOT CHANGE THIS
\usepackage[preprint]{aaai2027}  % DO NOT CHANGE THIS
\usepackage[hyphens]{url}  % DO NOT CHANGE THIS
\usepackage{graphicx} % DO NOT CHANGE THIS
\usepackage{natbib}  % DO NOT CHANGE THIS AND DO NOT ADD ANY OPTIONS TO IT
\usepackage{caption} % DO NOT CHANGE THIS AND DO NOT ADD ANY OPTIONS TO IT
\usepackage{booktabs}

\usepackage{amsmath}
\usepackage{amssymb}
\usepackage{multirow}
\usepackage{xcolor}
\usepackage{subcaption}
\usepackage{cleveref}  % AAAI-permitted; smart cross-references (\cref)
\crefname{figure}{Figure}{Figures}
\Crefname{figure}{Figure}{Figures}
\usepackage{tikz}
\usetikzlibrary{arrows.meta,positioning,calc,fit,backgrounds}
\definecolor{seqcol}{HTML}{C15A1B}    % sequence side (ESM-2 / prior sequence-SAE work)
\definecolor{structcol}{HTML}{147D6F} % structure side (folding trunk / this work)

\newcommand{\auroc}{\textsc{auroc}}

\newcommand{\ev}{\textsc{ev}}
\newcommand{\esmtwo}{ESM-2}
\newcommand{\esmfold}{ESMFold}
\newcommand{\withinaa}{within-AA}

\newcommand{\fitwidth}[1]{%
  \resizebox{\ifdim\width>\linewidth\linewidth\else\width\fi}{!}{#1}}

\newif\ifedited\editedtrue
\ifedited
  
\else
  
\fi

\crefname{appgeneric}{Appendix}{Appendices}
\Crefname{appgeneric}{Appendix}{Appendices}

\crefformat{appgeneric}{#2Appendix~#1#3}
\Crefformat{appgeneric}{#2Appendix~#1#3}
\crefmultiformat{appgeneric}{Appendices~#2#1#3}{ and~#2#1#3}{, #2#1#3}{, and~#2#1#3}
\Crefmultiformat{appgeneric}{Appendices~#2#1#3}{ and~#2#1#3}{, #2#1#3}{, and~#2#1#3}
\crefrangeformat{appgeneric}{Appendices~#3#1#4--#5#2#6}
\Crefrangeformat{appgeneric}{Appendices~#3#1#4--#5#2#6}

\title{Structure, Not Sequence: Tracking and Validating Folding Mechanisms in ESMFold's Trunk with Sparse Autoencoders}

\author{
    Bohan Liu\textsuperscript{\rm 1}\corresponding,
    Xiaohan Fan\textsuperscript{\rm 1},
    Wenqian Ye\textsuperscript{\rm 1},
    Aidong Zhang\textsuperscript{\rm 1}\corresponding
}
\affiliations{
    \textsuperscript{\rm 1}University of Virginia\\
    Charlottesville VA 22903, USA\\

    \{qzp4ta, aidong\}@virginia.edu
}
\begin{document}
\maketitle

\begin{abstract}
Structure predictors like \esmfold{} fold proteins using neural network computation and have
transformed structural biology. However, the driving mechanism from amino-acid (AA) sequence to 3D
structure is insufficiently understood, and biological meaning remains locked in the dense
network. Sparse autoencoders (SAEs) have shown promising results in obtaining interpretable concept
features over an overcomplete basis of intermediate protein sequence-model representations. Yet folding
models, which predict physical shape from AA sequence semantics, have no such account. We hypothesize that
the features carrying the folding computation evolve in the 3D structural regression process. In this paper, we interpret the folding trunk in three steps. First, a pilot analysis on the representation distribution across modules and recycles locates distinct spatial computation in the trunk. Second, the dictionaries expose a division of
labor: the single representation carries inherited semantics, and the pair representation
computes the pairwise geometry. Third, we interpret the dictionary at scale with a reasoning language model, trace
how the concepts compose across blocks by gradient-based attribution, and follow the
interpreted code to the model's folding output, verifying where the computation aligns with, or departs
from, known biochemistry. Together, the sparse dictionaries we train capture the relational geometry the
trunk computes beyond the sequence model, and their concept activations forecast the model's own fold quality, turning the black-box folding trunk into a sparse, interpretable, and scientifically verifiable circuit.
\end{abstract}

\section{Introduction}
\label{sec:intro}
\begin{figure}[t]
\centering
\resizebox{0.8\columnwidth}{!}{%
\begin{tikzpicture}[
font=\footnotesize,
stage/.style={rectangle, rounded corners=3pt, draw, line width=0.7pt, align=center,
inner sep=2.5pt, minimum width=32mm},
flow/.style={-{Stealth[length=2mm,width=1.6mm]}, line width=0.8pt},
slbl/.style={font=\footnotesize\itshape},               % stage sub-label
glbl/.style={font=\footnotesize},                       % glyph label
lossttl/.style={font=\scriptsize\bfseries},             % "X loss"
lossnote/.style={font=\scriptsize\itshape},             % loss sub-note
lead/.style={gray!60, line width=0.3pt}                 % plain example leaders
]

% --- Input sequence ---
\node (seq) [align=center]
{{\scriptsize\itshape amino-acid sequence}\\[1pt]{\small\ttfamily\bfseries\color{seqcol}MKVLSG{\ldots}HR}};

% --- ESM-2 sequence model ---
\node (esm) [stage, below=2.5mm of seq, fill=seqcol!10, draw=seqcol, minimum height=8.5mm]
{\textbf{\esmtwo{} Sequence Model}\\[-0.5pt]{\footnotesize\itshape\color{seqcol}sequence semantics}};

% --- Folding Trunk ---
\node (trunk) [stage, below=5.5mm of esm, fill=structcol!10, draw=structcol, line width=1.2pt,
minimum width=39mm, minimum height=21mm] {};

% --- Structure Module ---
\node (sm) [stage, below=5.5mm of trunk, fill=black!4, draw=black!45, minimum height=8.5mm]
{\textbf{Structure Module}};

% --- Output anchor ---
\node (out) [below=3mm of sm] {};

% --- Main Pipeline Flow Arrows ---
\draw[flow] (seq)--(esm);
\draw[flow] (esm)--(trunk);
\draw[flow] (trunk)--(sm);
\draw[flow] (sm)--(out.north);

% Trunk Title & Sub-label
\node[anchor=north, font=\bfseries] at ([yshift=-1.0mm]trunk.north) {Folding Trunk};
\node[anchor=north, slbl, text=structcol] at ([yshift=-4.8mm]trunk.north) {3D computation};

% s_s (1x5 single representation) & s_z (5x5 pair representation)
% Shifted North & perfectly vertically midpoint-aligned at y=9.0mm from trunk.south
\coordinate (ss_pos) at ([xshift=-10.25mm, yshift=8.25mm]trunk.south);
\begin{scope}[shift={(ss_pos)}]
\fill[structcol!20] (0,0) rectangle (7.5mm, 1.5mm);
\draw[step=1.5mm, black!45, line width=0.25pt] (0,0) grid (7.5mm, 1.5mm);
\draw[black!55, line width=0.4pt] (0,0) rectangle (7.5mm, 1.5mm);
\end{scope}

\coordinate (sz_pos) at ([xshift=2.75mm, yshift=5mm]trunk.south);
\begin{scope}[shift={(sz_pos)}]
\fill[structcol!30] (0,0) rectangle (7.5mm, 7.5mm);
\draw[step=1.5mm, black!45, line width=0.25pt] (0,0) grid (7.5mm, 7.5mm);
\draw[structcol, line width=0.5pt] (0,0) rectangle (7.5mm, 7.5mm);
\end{scope}

% Representation labels cleanly separated without collision
\node[glbl, anchor=north] at ([xshift=-6.5mm, yshift=5.5mm]trunk.south) {$s_s$ single};
\node[glbl, anchor=north] at ([xshift=6.5mm, yshift=5.5mm]trunk.south) {$s_z$ pair};

% Trained-on-structure background region
\begin{scope}[on background layer]
\node[fit=(trunk)(sm), inner sep=2.5mm, rounded corners=3pt,
fill=structcol!4, draw=structcol!35, dashed, line width=0.5pt] (trained) {};
\end{scope}

% Recycle loop & label
\draw[flow, structcol, line width=0.9pt]
(sm.east) -- ++(6.5mm,0) |- ([yshift=-4mm]trunk.north east);
\node[slbl, text=structcol, anchor=north west] at ([xshift=1mm,yshift=-0.8mm]sm.east) {recycle};

% Interpretation target label
\node (ours) [font=\footnotesize, text=structcol, align=center, left=4mm of trunk.west, anchor=east]
{\textbf{this work:}\\[-0.5pt]interpret $s_s,s_z$};
\draw[flow, structcol] (ours.east) -- (trunk.west);

% MLM loss (ESM-2 pretraining)
\coordinate (mlm) at ([xshift=-16mm]esm.west);
\begin{scope}[shift={(mlm)}]
\fill[seqcol!25,draw=seqcol,line width=0.3pt] (-4.5mm,0) rectangle ++(2.8mm,2.8mm);
\fill[seqcol,draw=seqcol,line width=0.3pt]    (-1.4mm,0) rectangle ++(2.8mm,2.8mm);
\fill[seqcol!25,draw=seqcol,line width=0.3pt] ( 1.7mm,0) rectangle ++(2.8mm,2.8mm);
\node[font=\scriptsize\bfseries,text=white] at (0mm,1.4mm) {?};
\draw[-{Stealth[length=1.4mm]},seqcol,line width=0.5pt] (0mm,3.2mm) -- ++(0,2.2mm);
\node[glbl,text=seqcol] at (0mm,6.3mm) {V};
\node[lossttl,text=seqcol] at (0mm,-2.5mm) {MLM loss};
\node[lossnote,text=seqcol] at (0mm,-5.0mm) {pretraining};
\end{scope}
\draw[-{Stealth[length=1.6mm]}, seqcol, dashed, line width=0.6pt] ([xshift=5.5mm]mlm) -- (esm.west);

% FAPE loss schematic
\coordinate (fape) at ([xshift=-19mm]sm.west);
\begin{scope}[shift={(fape)}]
% Reference graph (grey backbone + contact arc)
\draw[black!40, line width=0.8pt] (-7mm,1.8mm) -- (-2.5mm,4.5mm) -- (2.5mm,3.2mm) -- (7mm,1.5mm);
\draw[black!40, line width=0.8pt] (-7mm,1.8mm) to[bend left=28] (7mm,1.5mm);

% Predicted graph (teal backbone + contact arc)
\draw[structcol, line width=1.1pt] (-7mm,-1.6mm) -- (-2.5mm,1.6mm) -- (2.5mm,0.3mm) -- (7mm,-1.8mm);
\draw[structcol, line width=1.1pt] (-7mm,-1.6mm) to[bend left=28] (7mm,-1.8mm);

% Residual error vectors
\foreach \x/\yp/\yr in {-7/-1.6/1.8, -2.5/1.6/4.5, 2.5/0.3/3.2, 7/-1.8/1.5}
  \draw[{Stealth[length=1.0mm]}-, red!70!black, line width=0.5pt] (\x mm, \yp mm) -- (\x mm, \yr mm);
  
% Nodes
\foreach \x/\y in {-7/1.8, -2.5/4.5, 2.5/3.2, 7/1.5} 
  \fill[black!45] (\x mm, \y mm) circle (0.8mm);
\foreach \x/\y in {-7/-1.6, -2.5/1.6, 2.5/0.3, 7/-1.8} 
  \fill[structcol] (\x mm, \y mm) circle (0.8mm);
  
% FAPE Label
\node[lossttl, text=structcol] at (0mm, -4.5mm) {FAPE $+$ aux.\ loss};
\end{scope}
\draw[-{Stealth[length=1.6mm]}, structcol, dashed, line width=0.6pt] ([xshift=8mm]fape) -- (sm.west);

% --- Output visualizations ---
\coordinate (d3) at ([yshift=-2mm]out.south);
\begin{scope}[shift={(d3)}]
% Left side: 3D fold structure schematic curve
\begin{scope}[shift={(-12mm, 0)}]
\draw[structcol, line width=1.4pt, line cap=round]
(-6.5mm, 2mm) .. controls (-4.5mm, 5.5mm) and (-1.8mm, -2mm) .. (0mm, 2.5mm)
.. controls (1.8mm, 5.5mm) and (4.5mm, -1mm) .. (6.5mm, 2mm);
\node[anchor=north] at (0mm, -2.5mm) {{\scriptsize\itshape 3D fold structure}};
\end{scope}

% Right side: pLDDT confidence color map
\begin{scope}[shift={(11.5mm, 0)}]
  \shade[left color=blue!75!black, right color=orange!90!black] (-9mm, 1mm) rectangle (9mm, 3.2mm);
  \draw[black!40, line width=0.3pt] (-9mm, 1mm) rectangle (9mm, 3.2mm);
  \foreach \x/\v in {-9/0, 0/50, 9/100} 
    \node[font=\tiny, text=black!70, anchor=north] at (\x mm, 1.3mm) {\v};

  \node[anchor=north] at (0mm, -2.5mm) {{\scriptsize\itshape pLDDT confidence scale}};
\end{scope}
\end{scope}
\end{tikzpicture}}
\caption{Prior works interpret the \esmtwo{} sequence model trained on
masked-language modeling, whereas we read \esmfold{}'s folding trunk: its single ($s_s$) and pair
($s_z$) representations, supervised by the 3D frame-aligned (FAPE) structure loss together with
auxiliary read-out losses on the trunk states. The trunk and the structure head turn an amino-acid sequence embedding into a 3D structure with a
per-residue confidence (pLDDT), with optional iterative refinement denoted as \emph{recycle} in the figure. }
\label{fig:pipeline}
\end{figure}
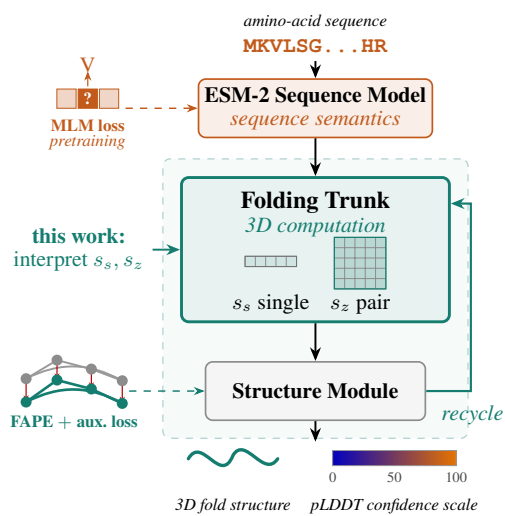
Deep neural structure predictors such as \esmfold{}~\citep{lin2023esm2} have revolutionized
structural biology by reducing computation times from days to seconds. Structure predictors transform a
protein's amino-acid (AA) sequence into a 3D structure through multiple learned
blocks. However, the internal computational processes remain opaque, rendering these folding models
black boxes that lack scientifically reliable explanations or guarantees. Verifying that the
computation aligns with underlying scientific processes, and extracting biological insight that links
hidden features to output structures, therefore remains out of reach.

Sparse autoencoders (SAEs), which have shown significant potential in interpreting the internal
computations of large language models (LLMs), are strong candidates for uncovering mechanisms underlying
protein structural regression. SAEs decompose dense activations into an overcomplete dictionary of
features that are substantially more interpretable than raw neurons \citep{bricken2023monosemanticity}.
In the context of proteins, SAEs have been trained on the sequence model \esmtwo{} to identify and
automatically interpret features that correspond to biological concepts
\citep{simon2024interplm,adams2025interprot}. However, because sequence language models are trained
on the masked language modeling objective, they only compute dense sequence embeddings and lack
explicit directives to learn information about the 3D structure that determines protein function. In contrast, the folding trunk transforms sequence embeddings into structural
representations. We therefore hypothesize that the trunk and the sequence model specialize in their distinct tasks: residue-level
semantics is inherited from the sequence embedding, while the geometry is computed inside the trunk and materialized by the structure module with optional refinement recycling. We visualize the \esmfold{} model design in \Cref{fig:pipeline}. 

Our work focuses on interpreting the folding trunk beyond the sequence model analyzed in
prior work, and emphasizes structure over sequence. Three motivating observations inform the approach:
folding hidden states are organized differently from the initial \esmtwo{} embeddings, the trunk
encodes structure in a pair representation that is absent from the sequence model, and trunk blocks compute spatial information across depth. These observations motivate a protocol, which involves training TopK SAEs at five evenly spaced trunk
blocks, automatically interpreting each resulting feature, and attributing features across blocks, detailed in \Cref{sec:method,sec:results}.
The resulting dictionary gives a diverse and faithful interpretation of the folding trunk. Measured on top of the language model states, the trunk code raises residue contact
and strand orientation detection by up to $+0.32$ \auroc{}, while residue-level properties are already
recoverable from the language model. Its features also localize functional sites, such as catalytic
residues and disulfide bonds, on the folded structure. Building upon the learned dictionary, a reasoning LLM assigns names
to each feature based on structural and functional context, while gradient-based attribution traces
the composition of these concepts across the trunk's depth into a directed folding circuit. Together, these steps give an automatically interpreted account of the transformation from sequence to structure.

The main contributions of this work are as follows:
\begin{itemize}
\item \textbf{Interpret the folding trunk at scale.} We train the first large-scale sparse-feature
dictionaries of a folding model's trunk module, on both its single and pair representations across
the depth of \esmfold{}, and interpret every feature with a reasoning LLM, verified by checking
each label on quantitative context withheld from the prompt assembler. 
\item \textbf{Locate the computation the sequence model leaves out.} We measure the trunk code's
contribution over the language-model states prior dictionaries decompose: gains of up to $+0.32$
\auroc{} on relations between residues, while residue-level properties are already supplied by the
language model. The pattern exposes a division of labor: the single representation carries inherited residue
semantics and the pair representation computes the geometry.
\item \textbf{Trace a causal folding circuit, and read the output from it.} We compose the concepts
into a directed circuit across depth with integrated-gradient attribution, confirm its edges causally
in the live \esmfold{} pass, trace the interactions between the single and pair representations within the trunk, and turn the interpretation into a read-out of fold quality.
\end{itemize}

\section{Related Work}
\label{sec:related}

\paragraph{Sparse autoencoders and mechanistic interpretability.}
SAEs decompose dense neural activations into a sparse, overcomplete dictionary of features that are
more monosemantic than raw neurons, a response to the superposition hypothesis under which networks
pack more features than they have dimensions~\citep{elhage2022superposition,bricken2023monosemanticity}.
They scale to frontier language models~\citep{templeton2024scaling,gao2025scaling} and admit
architectural variants: TopK SAEs that fix the active-feature count~\citep{gao2025scaling},
transcoders that predict one layer's activations from another~\citep{dunefsky2024transcoders}, and
crosscoders that share a single dictionary across layers or models~\citep{lindsey2024crosscoders}.
We build on this toolkit, training TopK dictionaries at five depths of the folding trunk. High
reconstruction and high single-feature selectivity still leave open whether a feature is causally used
or semantically meaningful~\citep{makelov2025towards}, and two families of tools address this. The
interventional family patches, ablates, or transplants activations and reads the downstream
effect~\citep{meng2022rome,wang2023ioi}; the correlational family, from the probing literature, tests a
probe against a control task solvable only by exploiting a confound, isolating whether the probe reads
the target property~\citep{hewitt2019control}. Our residue-identity control is the protein-model
instance of the latter, applied as attribution across a benchmark of interpreted features.

\paragraph{Protein models: structure, stability, and function.}
\esmtwo{} is a masked language model over protein sequences whose embeddings carry structural and
functional signal~\citep{rives2021biological,lin2023esm2}. \esmfold{} folds directly from these
embeddings through a trunk of structure blocks, in a single forward pass without an MSA
pipeline~\citep{jumper2021alphafold, lin2023esm2}. Beyond structure,
sequence likelihood and embedding features from the same models predict variant effects and
stability~\citep{meier2021language,brandes2023esm1b,frazer2021disease}, benchmarked at scale by
ProteinGym~\citep{notin2023proteingym}, and probing studies show that secondary structure, contacts,
and some functional properties are linearly decodable from Protein Language Model
embeddings~\citep{rao2019evaluating,vig2020bertology,rao2021transformer}. Single-sequence
predictors already reach high secondary-structure accuracy from sequence
alone~\citep{klausen2019netsurfp}, marking secondary structure as sequence-determined. The trunk
performs an explicit sequence-to-geometry regression, materializing 3D structure that
the frozen sequence embedding leaves implicit, and that computation is what we interpret. \Cref{fig:pipeline} places our target within the \esmfold{} pipeline.

\paragraph{Interpreting protein models.} On interpreting sequence semantics, 
InterPLM~\citep{simon2024interplm} introduced the language-model pipeline that
names protein-SAE features automatically, and InterProt~\citep{adams2025interprot} adds an interactive
latent viewer with human-rated interpretability; we adopt this automated-interpretation step directly.
On the structural side, \citet{parsan2025structuresae} train SAEs on the \esmtwo{}-3B sequence backbone
and relate the features to \esmfold{}'s output, leaving the folding trunk's internal computation and
residue-type correlation for future work. We interpret the \esmfold{} trunk directly, covering its single
representation $s_s$ and the folding-exclusive pair representation $s_z$ that no sequence-based
dictionary can reach, measure each feature by function, and separate what it adds from residue identity. We further trace interpreted concepts into a directed circuit across the trunk's
blocks with integrated gradients, and close the gap between deep folding networks and interpretable biological processes. \looseness=-1

\section{Method}
\label{sec:method}

Our protocol trains sparse dictionaries on the internal states of \esmfold{}'s folding trunk,
interprets each resulting feature with a language model, evaluates what the features contribute over
the sequence model, and traces how features compose across the trunk's depth.
\Cref{sec:prelim-study} defines the trunk
representations and the three observations that motivate the protocol, \Cref{sec:dict} gives the
dictionary recipe, \Cref{sec:autointerp} describes the automated interpretation step we adopt from prior
protein-SAE work, \Cref{sec:eval} defines the evaluation strategy,
and \Cref{sec:circuit} describes the integrated-gradient construction of the folding circuit. Corpus, optimization, and protocol details are collected in \Cref{sec:impl}.

\subsection{The folding trunk and pilot analysis}
\label{sec:prelim-study}

\paragraph{Trunk representations.}
\esmfold{}~\citep{lin2023esm2} folds a sequence by first embedding it with the \esmtwo{}-3B
masked language model~\citep{lin2023esm2} and then passing the embedding through a folding trunk of
$48$ structure blocks. Each block maintains two states: a \emph{single representation}
$s_s\in\mathbb{R}^{L\times 1024}$, one vector per residue, and a \emph{pair representation}
$s_z\in\mathbb{R}^{L\times L\times 128}$, one vector per residue pair, which has no counterpart in
the sequence model. The trunk is applied \emph{iteratively}: at each of a fixed number of recycles a
structure module reads $s_s$ and $s_z$ into a predicted structure, and the next iteration restarts the
trunk with both representations carried over and the predicted structure returned as a binned
distogram. The coordinates from the final recycle are the output. We analyze five evenly spaced blocks, $b_{00},b_{12},b_{24},b_{36},b_{47}$, spanning
the trunk from its first block ($b_{00}$) to the block that feeds the
structure module ($b_{47}$).

\begin{figure*}[t]
\centering
\includegraphics[width=0.8\textwidth]{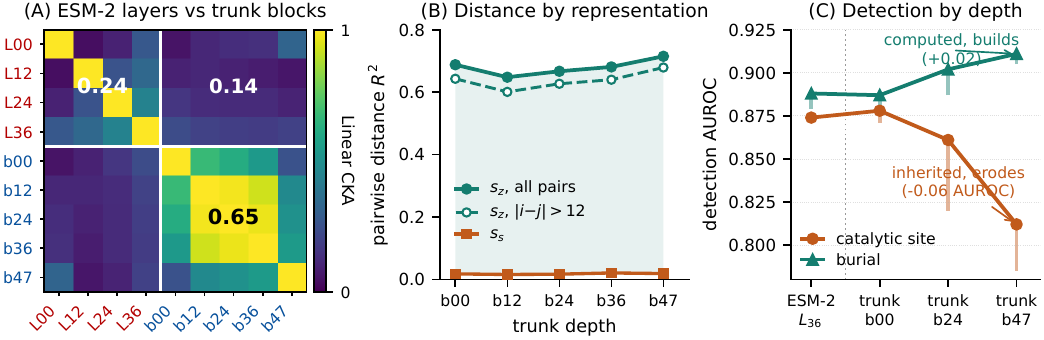}
\caption{\textbf{Three pilot observations motivate the protocol.} \textbf{(A)}~Linear CKA between \esmtwo{} layers and trunk blocks. \textbf{(B)}~Distance $R^2$ for pairwise residue geometry decoded from the pair representation $s_z$ versus any single-residue state $s_s$, by sequence separation. \textbf{(C)}~Detection \auroc{} across trunk depth, from the frozen \esmtwo{} $L_{36}$ to $b_{47}$, for a functional-site concept (catalytic) and a geometric one (burial).}
\label{fig:prelim}
\end{figure*}

\paragraph{Empirical motivation.}
Three properties of the trunk shape our protocol and analysis:
\emph{(i)~A representational gap.} We measure linear Centered Kernel Alignment~(CKA)~\citep{pmlr-v97-kornblith19a} between the \esmtwo{}
layers and the trunk blocks, plotted in \Cref{fig:prelim}A. \esmtwo{}--trunk block pairs average $0.14$, and \esmtwo{}'s final layer
 reaches $0.23$ CKA with $b_{00}$; the trunk blocks meanwhile
agree with one another at averaged $0.65$, and the middle blocks peak at $0.96$. The trunk organizes residues by a
similarity structure of its own, which calls for dictionaries trained on that structure rather than
the sequence dictionaries of prior work~\citep{simon2024interplm,adams2025interprot}.
\emph{(ii)~Structure in the pair representation.} Geometry is encoded in the folding-exclusive pair
representation $s_z$: pairwise residue distance is linearly decodable from $s_z$ at deep blocks and
stays near-absent from any single-residue state, as \Cref{fig:prelim}B shows, so a faithful account of the
folding computation must read $s_z$ as well as $s_s$. \emph{(iii)~Inherited residue semantics, computed geometry.} A depth-resolved probe shows that functional-site signal is maximal at \esmtwo{} $L_{36}$ and
erodes with trunk depth, plotted in \Cref{fig:prelim}C. The early-commitment pattern indicates that per-residue
semantics is inherited from the
sequence model, whereas geometric signal such as burial builds with depth and
is therefore computed inside the trunk. We cache cycle-1 and cycle-3 latents for training. \Cref{app:recycle} shows the representation is insensitive to the recycle budget. This divergence between functional and geometrical information along the trunk depth motivates the interpretation of the within-trunk computation.

\subsection{Sparse dictionaries of the trunk}
\label{sec:dict}

\paragraph{TopK dictionaries.}
For each block we train a TopK sparse
autoencoder~\citep{gao2025scaling,bricken2023monosemanticity} on the block's
representation, mean-centered over the training corpus. A residue (or
pair) activation $x$ is encoded and reconstructed as
\begin{equation}
z = \mathrm{TopK}_k\!\big(\mathrm{ReLU}(xW_{\mathrm{enc}} + b_{\mathrm{enc}})\big),
\hat{x} = z W_{\mathrm{dec}} + b_{\mathrm{dec}},
\label{eq:sae}
\end{equation}
where $\mathrm{TopK}_k$ keeps the $k$ largest ReLU logits and zeros the rest. Training minimizes reconstruction error
$\lVert x-\hat{x}\rVert_2^2$ with the decoder rows held at unit norm and dead features periodically
reinitialized toward high-residual examples. We report explained variance (\ev{}) on held-out data.

\paragraph{Where dictionaries are placed.}
We train a dictionary at each of the five trunk blocks on both the single representation $s_s$ and the
pair representation $s_z$, and, as a sequence baseline, on the \esmtwo{} layers that feed the trunk. We set $k{=}16$ throughout to obtain balanced reconstruction
and sparsity for interpretation. The training corpus, optimization budget, and checkpoint format are detailed in
\Cref{sec:impl}.

\subsection{Automated interpretation}
\label{sec:autointerp}

The two representations carry different objects, so we interpret them with two prompts. For the
single representation we start from the pipeline introduced for protein SAEs by
InterPLM~\citep{simon2024interplm}: a reasoning LLM reads a context assembled from the
feature's top-activating residues across proteins with the structural and functional environment at
each, and returns a short free-form label and one of seven concept categories: secondary structure, burial, contact,
functional site, disulfide, chemical, positional. The reasoning LLM also decides whether the feature reflects
amino-acid type or a real biological concept.

A pair feature operates on residue pairs. As prior interpretations missed this object, our revised prompt
describes the population of pairs a feature fires on and gives both endpoints' residue context. The prompt asks openly what
relation the pair encodes. Because the representations are interpreted differently, we validate the distinction between representations with cross-prompt control in \Cref{app:crossprompt}. \looseness=-1

\subsection{Feature evaluation}
\label{sec:eval}

\paragraph{Feature localization on the protein.}
We localize feature $f$ by its activation sites, the map
\begin{equation}
a_i = z_{i,f},
\label{eq:gradcam}
\end{equation}
painted on the folded structure. The encoder applies per residue, so this activation is the
feature's complete read-out of each residue: a gradient
$\partial\big(\textstyle\sum_j z_{j,f}\big)\big/\partial s_{s,i}$ is the encoder row on the
feature's support and vanishes off it.

\paragraph{Isolating the trunk's contribution from the sequence model.}
\label{sec:relational}
To isolate the additive trunk computation, we measure the trunk code as an increment over the sequence latents. For
each target we fit a probe on \esmtwo{}-3B at $L_{36}$, its final layer, which the trunk's learned
mixture over the \esmtwo{} representations weights at $0.90$, then on those
same states concatenated with the trunk code, and read the difference in held-out \auroc{} across
proteins. Targets come in two forms: per-residue properties, namely annotated sites, burial and
strand, and relations between two residues, namely residue contact, $\beta$-ladder continuation and
parallel-versus-antiparallel strand orientation. On a relational target every representation is given the pairwise interaction
features $[\,s_i;\,s_j;\,s_i\!\odot\! s_j\,]$, so a per-residue state is asked about the pair
through the same probe the trunk code gets. Strand orientation is scored within a common distance shell, so proximity cannot stand in for topology. 

\paragraph{Relating the code to the folding output.}
We fit a ridge probe from a block's code to both \esmfold{}'s own confidence
(pLDDT) and accuracy against the ground truth (lDDT). We report held-out $R^2$ on
unseen proteins. Individual concepts are then ranked by their Spearman correlation with accuracy. 

\subsection{Circuit construction by integrated gradients}
\label{sec:circuit}

To trace how interpreted concepts compose, we attribute each downstream feature to the upstream code
by integrated gradients~\citep{sundararajan2017ig} through the live \esmfold{} forward pass. One
construction covers every attribution we report, parameterized by a source block, a destination block,
and a choice of representation at each end. At the source we splice in the dictionary of the chosen
representation; at the destination we capture the chosen representation and encode it with its own
dictionary. The edge from upstream feature $i$ to downstream
feature $j$ is
\begin{equation}
E_{ij} = z_i \int_0^1
  \frac{\partial z^{\mathrm{dst}}_j\!\big(\alpha z\big)}{\partial z_i}\,\mathrm{d}\alpha,
\label{eq:edge}
\end{equation}
the integrated gradient of the downstream code along the straight path from the zero-code baseline to
the realized upstream code $z$. Each attribution runs on a single recycle, as the change across recycles is
minimal.

The four combinations give the whole circuit over blocks and representations. Within-domain tracing gives the information flow within a representation track, $s_s\!\to\!s_s$
through the single representation and $s_z\!\to\!s_z$ through the pair representation. Cross-attribution
traces the coupling between the representations, forward as $s_s\!\to\!s_z$ where the single
representation writes into the pair, and backward as $s_z\!\to\!s_s$ where the pair representation
writes back. \Cref{app:ig} provides details. \looseness=-1

% \paragraph{Causal test.}
% An attributed edge is a claim about routing, which we test causally in the native forward pass: we
% ablate or steer the attributed upstream feature and measure the change in its downstream partner
% against a control matched on both magnitude and direction. This distinguishes a genuine
% feature-to-feature edge from a shift that any perturbation of comparable size would produce, and it
% supports the composition claims of \Cref{sec:res-circuits} and the output-attribution claims of
% \Cref{sec:res-output}.

\section{Experiments and Results}
\label{sec:results}

\subsection{Implementation details}
\label{sec:impl}

\paragraph{Dictionary training.}
The single-representation cache is created from
$500$k folded UniRef50 sequences, giving a corpus of $95$M single-residue activations. In order to make fair trunk-versus-sequence comparisons, every single-representation and \esmtwo{}-layer dictionary is trained on the same data budget and the following protocol: batch size $bs=2048$, $k{=}16$, dictionary width $d_{\mathrm{sae}}{=}8192$, number of epochs $ep=10$, a linear
learning-rate warmup over the first $5\%$ of steps, and base learning rate
$lr = 4\times10^{-4}$. Pair-representation dictionaries use \eqref{eq:sae} with
$d_{\mathrm{in}}{=}128$ and $d_{\mathrm{sae}}{=}4096$. Because the number of pair samples grows with
$L^2$, they are trained under a different recipe: $4.8$M pair activations drawn from $967$ proteins,
batch size $8192$, twenty-five epochs. Dictionaries are trained at the five trunk blocks
and at the \esmtwo{} layers the trunk reads from. A sparsity sweep over
$k\in\{16,32,64,128\}$ is provided in \Cref{tab:ksweep}.

\paragraph{Interpretation context.}
For every feature, the reasoning LLM's context is assembled from its top-activating residues
across proteins, annotated with the structural and functional environment: secondary structure
and solvent-accessible area from the AlphaFold model~\citep{varadi2022alphafolddb,kabsch1983dssp,tien2013maximum},
the number of $\mathrm{C}\alpha$ contacts and long-range tertiary neighbors, disulfide participation, and
the UniProt functional annotation at that position~\citep{uniprot2023}. Alongside each label we record the
feature's amino-acid purity and dominant residue as a quantitative check on the identity-versus-concept
decision. A pair-representation feature is instead described by the distribution over its activating
pairs, and its question asks openly what relation the pair encodes. The pipeline yields an interpreted
label for $5{,}595$ alive single-representation features and $2{,}972$ pair features across the trunk
blocks. \Cref{app:prompts} reproduces both prompt templates.

\paragraph{Evaluation protocol.}
The increment over the sequence model of \Cref{sec:res-relational} samples positions from each protein with equal weights: $70$ residues per
protein where the target is a per-residue property and $220$ pairs per protein where it is a relation,
drawn without replacement under a fixed seed. Its probe is a logistic regression under a
protein-stratified split, so no protein contributes to both training and evaluation.

Reported confidence intervals are $95\%$ nonparametric bootstraps, randomly resampled as specified in
\Cref{app:ci}.

\subsubsection{Dictionary evaluation}
\label{sec:res-recon}
TopK dictionaries trained on the trunk's single representation reconstruct it well across depth:
at the $k{=}16$ setting we use throughout, explained variance is \ev{}$=0.87,0.90,0.89,0.86,0.89$ for blocks $0,12,24,36,47$, trading a few points of reconstruction
for interpretable codes. 

\begin{figure*}[t]
\centering
\includegraphics[width=0.88\textwidth]{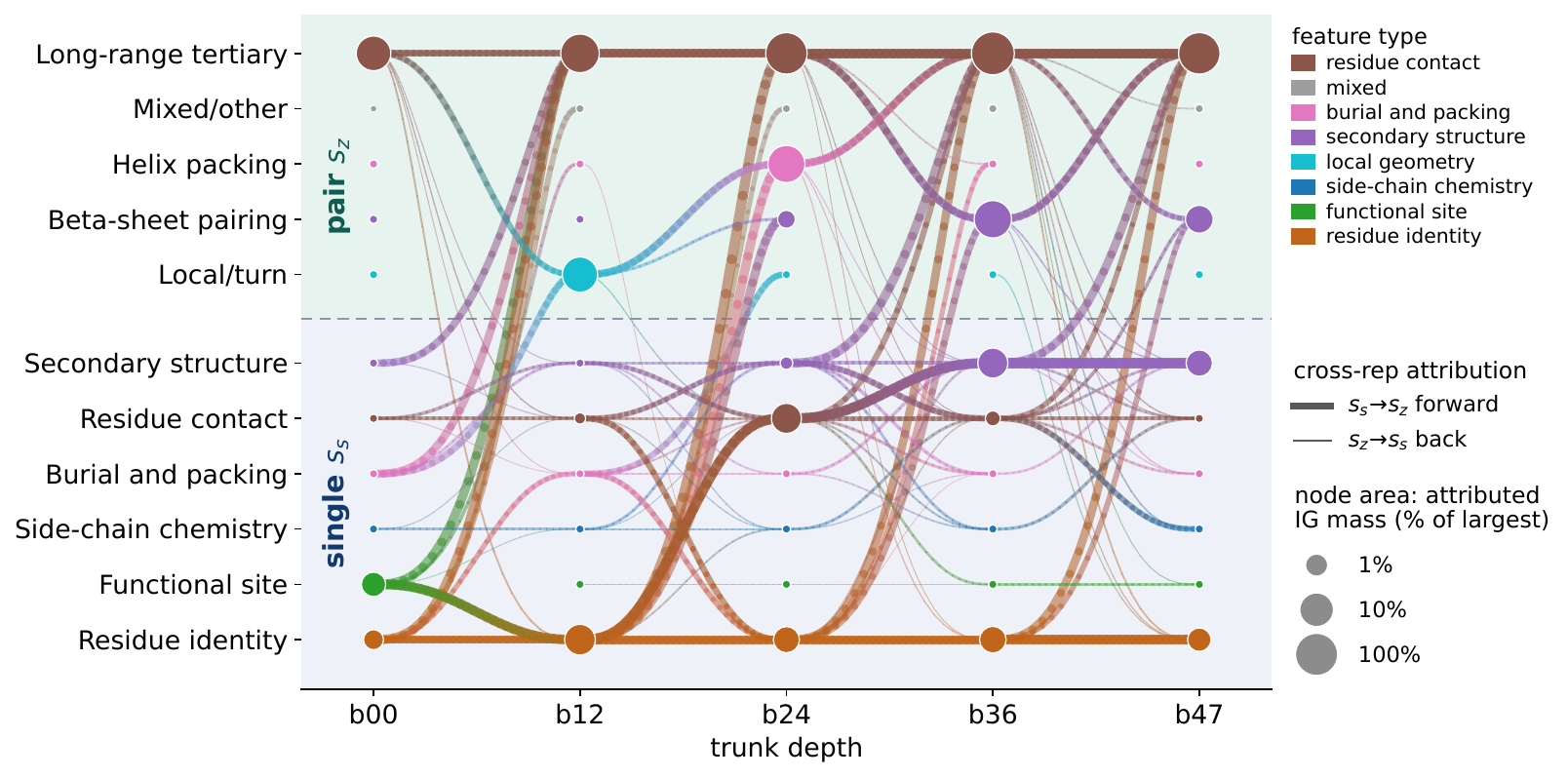}
\caption{The interpreted circuit among the annotation categories. The categories are the nodes, and
integrated gradients route between them across the trunk's depth. The pair representation $s_z$ (top
track) carries the geometry classes, the single representation $s_s$ (bottom track) the residue-level
categories. Two cross-track bundles carry the coupling on a shared width scale, with forward
$s_s\!\to\!s_z$ thick and backward $s_z\!\to\!s_s$ thin.}
\label{fig:conceptflow}
\end{figure*}

\subsection{Growing relational information in the trunk}
\label{sec:res-relational}
We measure what the trunk code supplies on top of the per-residue language-model states that prior
dictionaries decompose, reading the \auroc{} increment when the trunk code is added to the
\esmtwo{}-3B $L_{36}$ probe defined in \Cref{sec:relational}. Because that baseline already encodes sequence semantics,
the increment isolates what the trunk adds beyond semantics. On relations between residues, adding the trunk code improves prediction substantially. Residue contact gains $+0.318$ \auroc{}, ladder continuation $+0.209$, and strand
orientation $+0.086$, each far outside the spread of a control that shuffles the trunk features across
examples ($\pm 0.010$), as \Cref{tab:relational} reports.\looseness=-1

Strand orientation is the cleanest of these. Among residue pairs that are already paired, whether the
two strands run parallel or antiparallel is settled by topology alone: the two classes differ in mean
separation by $0.05$\,\AA{}, and distance by itself scores $0.511$. The trunk pair code reaches
$0.965$, where \esmtwo{}-3B under the same pairwise probe reaches $0.862$.

Residue-level properties behave differently. Active site gains $+0.006$, ligand binding $+0.001$,
burial $-0.000$, and strand $+0.008$, all within the shuffled control, with metal binding marginally
above it at $+0.014$. The
language model already supplies what a per-residue read-out can use for these, and the pattern follows
the form of the target rather than its difficulty: burial leaves ample headroom at $0.827$ and still
gains nothing. The same boundary appears at the whole-protein level, where the pooled trunk code
matches an \esmtwo{}-3B dictionary at enzyme-class and fold retrieval, as \Cref{app:retrieval} reports.

\begin{table}[t]
\centering\small
\setlength{\tabcolsep}{4pt}
\resizebox{0.9\columnwidth}{!}{
\begin{tabular}{llccc}
\toprule
Target & Type & \esmtwo{}-3B & $+$ trunk & $\Delta$ \\
\midrule
Residue contact      & relational   & 0.636 & 0.954 & \bf +0.318 \\
Ladder continuation  & relational   & 0.644 & 0.853 & \bf +0.209 \\
Strand orientation   & relational   & 0.873 & 0.959 & \bf +0.086 \\
\midrule
Metal binding        & per-residue  & 0.978 & 0.992 & +0.014 \\
Strand               & per-residue  & 0.961 & 0.969 & +0.008 \\
Active site          & per-residue  & 0.978 & 0.984 & +0.006 \\
Ligand binding       & per-residue  & 0.923 & 0.924 & +0.001 \\
Burial               & per-residue  & 0.827 & 0.827 & $-$0.000 \\
\bottomrule
\end{tabular}
}
\caption{What the trunk code adds to the language-model states prior dictionaries are trained on.
\textbf{Target}: the property a probe is asked to predict. \textbf{Type}: whether the target is a
relation between two residues or a property of one. \textbf{\esmtwo{}-3B}: \auroc{} of a probe on
\esmtwo{}-3B $L_{36}$ alone, with the pairwise probe $[\,s_i;\,s_j;\,s_i\!\odot\! s_j\,]$ for
relational targets. \textbf{$+$ trunk}: the same states concatenated with the trunk code.
\textbf{$\Delta$}: the increment, to be read against a control that shuffles the trunk features across
examples, which spans $\pm 0.010$ on every row.}
\label{tab:relational}
\end{table}

\subsection{Faithful automated interpretation of concepts}
\label{sec:res-autointerp}
The pipeline of \Cref{sec:autointerp} labels $8{,}567$ alive features across the five blocks, $5{,}595$
in the single representation and $2{,}972$ in the pair representation, giving each a free-form label, a
concept category, and whether it is a residue-type indicator. The single-representation labels recover
quantities withheld from the prompt: amino-acid purity predicts a feature's identity assignment at
\auroc{}$=0.85$ (mean purity $0.85$ for identity-labeled features, $0.45$ for structural
ones), and the annotation density of its top contexts predicts its functional-site assignment at $0.89$
($0.59$ to $0.14$). Of the $3{,}502$ single-representation features carrying a spatial concept,
$1{,}270$ name packing, burial, or tertiary contact with no annotation anywhere in their context, so the vocabulary describes geometry the trunk computes, without echoed annotation. We separate the sequence-oriented categories from the 3D-oriented ones and observe a distinct separation between the two tracks of single and pair representations. The pair vocabulary is almost entirely geometric,
$91\%$ geometric to $9\%$ sequence, while the single vocabulary is mixed, $57\%$ geometric to $43\%$
sequence, carrying five times the pair track's share of sequence concepts alongside geometry information. Within the single representation the geometry share falls with depth, from $0.75$ at $b_{00}$ to
$0.43$ at $b_{47}$, so sequence semantics concentrates at the read-out. To test for bias induced by the heterogeneous automated interpretation, we perform a cross-prompt control experiment. When asked about sequence semantics, the pair vocabulary
stays geometric at $99\%$, and when asked about spatial relations, the single vocabulary does not turn more
geometric, falling from $57\%$ geometric share to $32\%$, as \Cref{fig:autointerp}A reports. To validate the interpreted concepts, we probe the pair representation's geometric concepts on relations read from coordinates, and each group predicts its own relation: $\beta$-labeled features reach \auroc{} $0.938$ on strand pairing against $0.839$ on the helix
contact, while helix-labeled features reverse the ordering, $0.989$ on the helix contact against $0.737$ on strand pairing. Both clear a random group of the same size, as \Cref{app:labelfirst} reports. The
category profile across depth and the assignment-grounding curves are provided in \Cref{fig:autointerp}.

Clustering the free-form labels blind to the taxonomy, with TF-IDF and $k$-means, recovers a
data-driven vocabulary, and \Cref{fig:conceptflow} traces how its concepts compose into a circuit across depth.
The pair representation yields $18$ concepts that are entirely geometric relations: residue contact,
$\beta$-strand pairing, helix periodicity, and hydrophobic packing, with no residue-identity concept;
the single representation yields $28$ concepts that add a near-complete single-residue lexicon ($13$ of
$28$, one per amino acid) concentrating at the read-out. The concepts are grounded in measurements
independent of the label text: identity concepts name the measured
dominant residue in $12$ of $13$ clusters with $0.84$ purity, above the baseline $0.40$ of structural concepts. Geometry concepts match the measured sequence separation, $\sim\!6$ residues for helix periodicity compared to $\sim\!36$ for long-range and tertiary contact.

\subsection{Activation sites}
\label{sec:res-reading}

\begin{figure}[t]
\centering
\includegraphics[width=0.8\columnwidth]{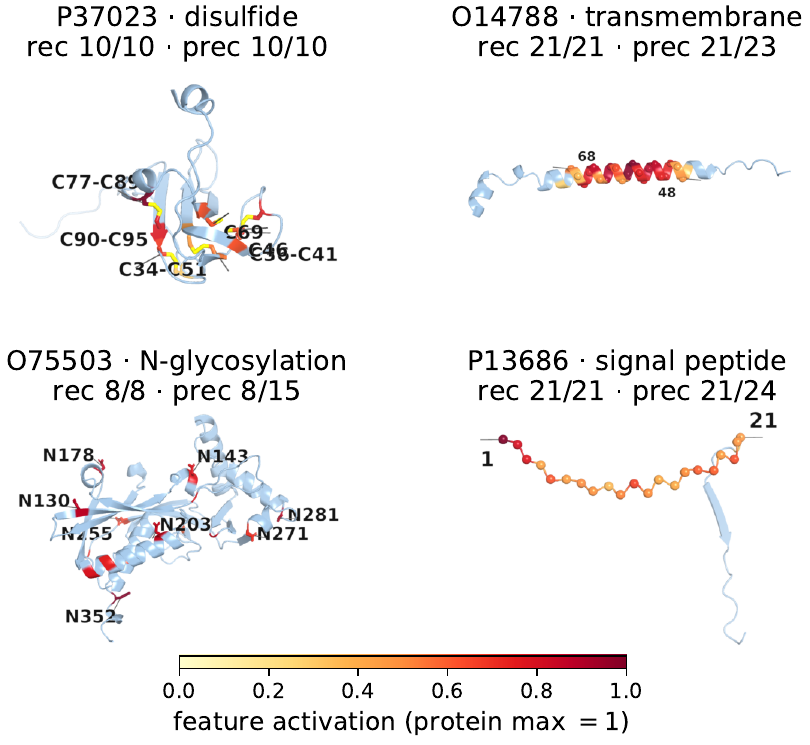}
\caption{Activation sites of select concepts and proteins. The activation is clipped below 0.15 for clarity. Within-protein recall and precision are denoted as rec and prec respectively. }
\label{fig:gradcam}
\end{figure}

The quantitative account rests on a dictionary that can be inspected directly, feature by feature, with
labels that, as \Cref{sec:res-autointerp} establishes, answer to context the reasoning LLM never
sees. For a detector
already chosen by the detection scan, the activation sites of
\eqref{eq:gradcam} localizes on the folded structure the residues that the feature reads from.
\Cref{fig:gradcam} shows this across four annotation types: the disulfide detector concentrates on the
bonded cysteine pairs, and the transmembrane, signal-peptide, and N-glycosylation detectors on their
annotated spans, with more examples in \Cref{fig:gradcam_full}. Out of sample, the same detectors annotate
proteins carrying no functional-site label, and each annotation is a claim about coordinates that can
be tested without consulting the detector. Of the cysteines flagged as disulfide-bonded across $201$
such proteins, $594$ of $1{,}121$ take a C$\beta$--C$\beta$ partner within $6.5$\,\AA{} in the
predicted fold. \Cref{app:apps} reports those annotations, their sequence-level counterparts, and a
panel of disease enzymes.
\begin{table*}[!t]
\centering\small
\setlength{\tabcolsep}{3.5pt}
\resizebox{0.95\textwidth}{!}{
\begin{tabular}{llll}
\toprule
Edge & Source feature & Target feature & Explanation \\
\midrule
\multicolumn{4}{l}{\emph{supported by known biochemistry}}\\
$b_{12}\!\to\!b_{24}$, fwd & Leucine & Long-range ${\sim}12$\,\AA{} C$\beta$--C$\beta$ & leucine tends to bury, and buried residues meet at contact distance \\
$b_{24}\!\to\!b_{36}$, fwd & Exposed acidic loop residue & Long-range ${\sim}10$\,\AA{} proximity & acidic residues tend to be exposed, and exposed residues tend to pair \\
$b_{00}\!\to\!b_{12}$, fwd & Active sites or termini & Long-range tertiary contact & a catalytic site requires residues far apart in sequence to meet \\
$b_{00}\!\to\!b_{12}$, fwd & Glycine residue identity & Glycine-mediated contact & the target names glycine, the residue the source detects \\
\midrule
\multicolumn{4}{l}{\emph{rejected}}\\
$b_{36}\!\to\!b_{47}$, fwd & Helix termini and tight turns & Long-range exposed-coil contact & a turn constrains the backbone locally, the target spans the fold \\
$b_{00}\!\to\!b_{12}$, fwd & Loop/coil/turn residue & Long-range ${\sim}10$\,\AA{} contacts & a loop constrains the backbone locally, the target spans the fold \\
$b_{36}\!\to\!b_{47}$, bwd & Alpha-helix $i,i{+}4$ contact & Valine & valine is $\beta$-branched and disfavors helix \\
$b_{12}\!\to\!b_{24}$, fwd & Glycine and Histidine residues & Cysteine disulfide pairs & a disulfide requires cysteine, which neither residue supplies \\
\bottomrule
\end{tabular}
}
\centering
\caption{The eight chosen attribution cases. \textbf{Edge}: the attribution edge location, and whether it runs forward
($s_s\!\to\!s_z$) or backward ($s_z\!\to\!s_s$). \textbf{Source} and \textbf{Target}: the auto-interpreted features. \textbf{Explanation}: the biochemical explanation.}

\label{tab:cases}
\end{table*}

\subsection{Direction of information flow}
\label{sec:res-circuits}

The interpreted concepts of \Cref{sec:autointerp} compose into a directed circuit across the trunk's
depth, drawn in \Cref{fig:conceptflow} by attributing each downstream feature to the
upstream code with integrated gradients. Row names are display
forms picked for clarity, with their full details in \Cref{app:flowrender}.
The single representation is seeded from the sequence embedding and the pair
representation holds the geometry the fold is made of, so we call $s_s\!\to\!s_z$ the forward
direction, from sequence toward structure, and $s_z\!\to\!s_s$ the backward one. Attributed edges claim that one interpreted feature drives another, and
known biochemistry supports most of them: $79\%$ of edges within either representation, $75\%$ forward and $63\%$ backward across them. \Cref{app:igfull} draws each of the four attributions in
full at the level of the individual labels. Attributed edges are also tested in the live
\esmfold{} pass, by ablating the upstream feature at its block and reading the change in its
attributed downstream partner against a write matched on both magnitude and direction. The edge moves
its partner $1.42\times$ that control; \Cref{app:controls,app:szcausal} report the
pair-representation effects and the control variants. \looseness=-1

The two representations flow unequally: about a third of the pair code's attribution traces to
$s_s$, led by hydrophobic identity and core burial, while the geometry feeds back into the
single-representation code only weakly, about $6\%$.

\subsection{Case study: biochemically supported and rejected edges}
\label{sec:res-cases}

\Cref{tab:cases} traces the most strongly attributed edge of each kind, so the cases show what a
verdict is made of rather than how the verdicts are distributed. The supported edges are the ones
coherent with biochemistry. The rejected ones diverge in their causes: two carry a local cause into
long-range geometry, while the other two bind a residue to geometry against its chemical property. \looseness=-1

Across all rejections, most fall on the backward direction: $37\%$ of backward
cross-track edges are unsupported, compared to $25\%$ of forward ones, with $248$ out of the $290$ rejected
cross-track edges being backward. Because forward edges carry the larger attributed weight, the trend is not reflected in \Cref{tab:cases}.

\subsection{Influence on the folding output}
\label{sec:res-output}
\begin{figure}[!t]
\centering
\includegraphics[width=\columnwidth]{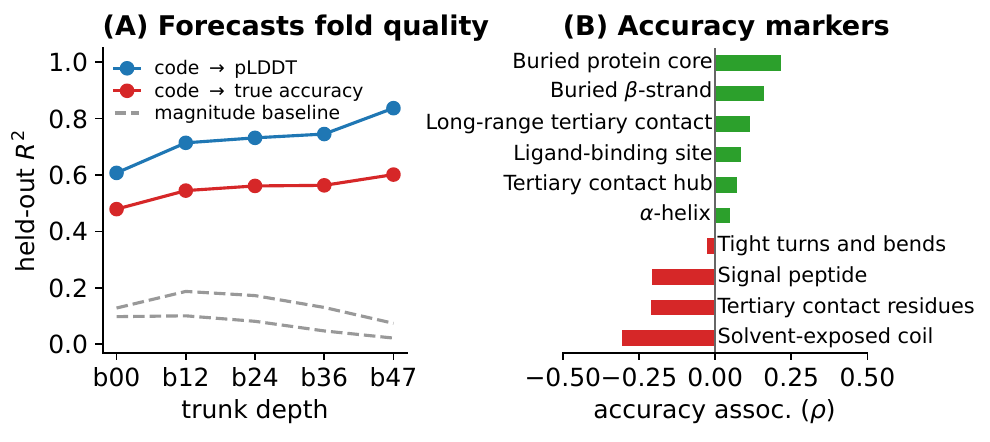}
\caption{Interpreted features forecast fold quality ($n{=}70{,}859$ residues, $452$ proteins).
\textbf{(A)}~A ridge probe on the interpretable code predicts \esmfold{}'s per-residue confidence
(pLDDT) and its true per-residue accuracy (lDDT against the reference structure) at every trunk
block.
\textbf{(B)}~Structural concepts ranked by their association with true accuracy (Spearman $\rho$); green marks positive and red marks negative correlation.}
\label{fig:output}
\end{figure}

Beyond describing the trunk, the code forecasts how well it folds. \Cref{fig:output}A plots a ridge
probe on the interpretable code predicting \esmfold{}'s per-residue confidence (pLDDT) and its true
per-residue accuracy (lDDT against the reference structure) at every trunk block, rising from
$R^2{=}0.61$ and $0.48$ at the first block to $R^2{=}0.84$ and $0.60$ at the read-out and
standing far above a representation-magnitude baseline. \Cref{fig:output}B ranks the structural concepts by accuracy: the buried and packed ones (buried
core, buried $\beta$-strand, long-range tertiary contact) mark accurate residues, while solvent-exposed, flexible
and terminal concepts mark less confident and less accurate ones.

\section{Conclusion}
\label{sec:conclusion}
We open the black box of \esmfold{}'s folding trunk and interpret it into a circuit of named concepts. Our trained sparse dictionaries over both tracks of the
trunk representations, interpreted at scale by a reasoning LLM, create a rich and faithful interpretation starting point.
We observe the split of task between the representation tracks: the pair representation is $91\%$ geometric, and the single representation keeps the sequence
semantics intact from the sequence language model. Measured on top of the language-model embedding,
the trunk code adds up to $+0.32$ \auroc{} to inferring residue relational properties.
Integrated gradients compose the concepts into a directed circuit, with known biochemistry supporting about
$73\%$ of its edges. We hope this work can promote more responsible application of the deep folding models and inspire future work on interpretable AI for Biochemistry. \looseness=-1

\bibliography{references}

\clearpage
\appendix

% Alias only the unnumbered appendix SECTIONS to the generic "Appendix" string. Figures and tables
% keep their own numbered types: aliasing them made every \Cref{fig:...}/\Cref{tab:...} in (and into)
% the appendix render as "Appendix" -- e.g. "Appendix profiles the fixed-taxonomy vocabulary" and the
% mashed "AppendixA" for a panel-letter reference.
\crefalias{section}{appgeneric}
\crefalias{subsection}{appgeneric}

% \section{Scope and limitations}
% \label{app:limits}
% Two parts of \esmfold{} sit outside what we interpret. The trunk carries four supervised read-outs,
% and we treat two of them: the distogram head over the pair representation and the language head over
% the single one. A pTM head also reads the pair representation and yields the predicted aligned error,
% a pairwise quantity, and the confidence head reads the structure module's own states rather than the
% trunk. Our claim that a relation escapes every read-out is therefore established for the specific
% relations we test, strand orientation among them, rather than for pairwise quantities in general, and
% extending the dictionaries to the structure module's states is left to future work.

% The single representation also receives a learned per-residue embedding of the amino acid alongside
% the sequence-model projection, so residue identity reaches the trunk by a second path. Where we
% attribute identity concepts to inheritance from the sequence model, that inheritance is from the
% sequence pathway as a whole rather than from \esmtwo{} alone.
\section*{\LARGE Appendix}

\section{Supplementary interpretability figures}
\label{app:suppfigs}

\begin{figure*}[p]
\centering
\includegraphics[height=0.92\textheight]{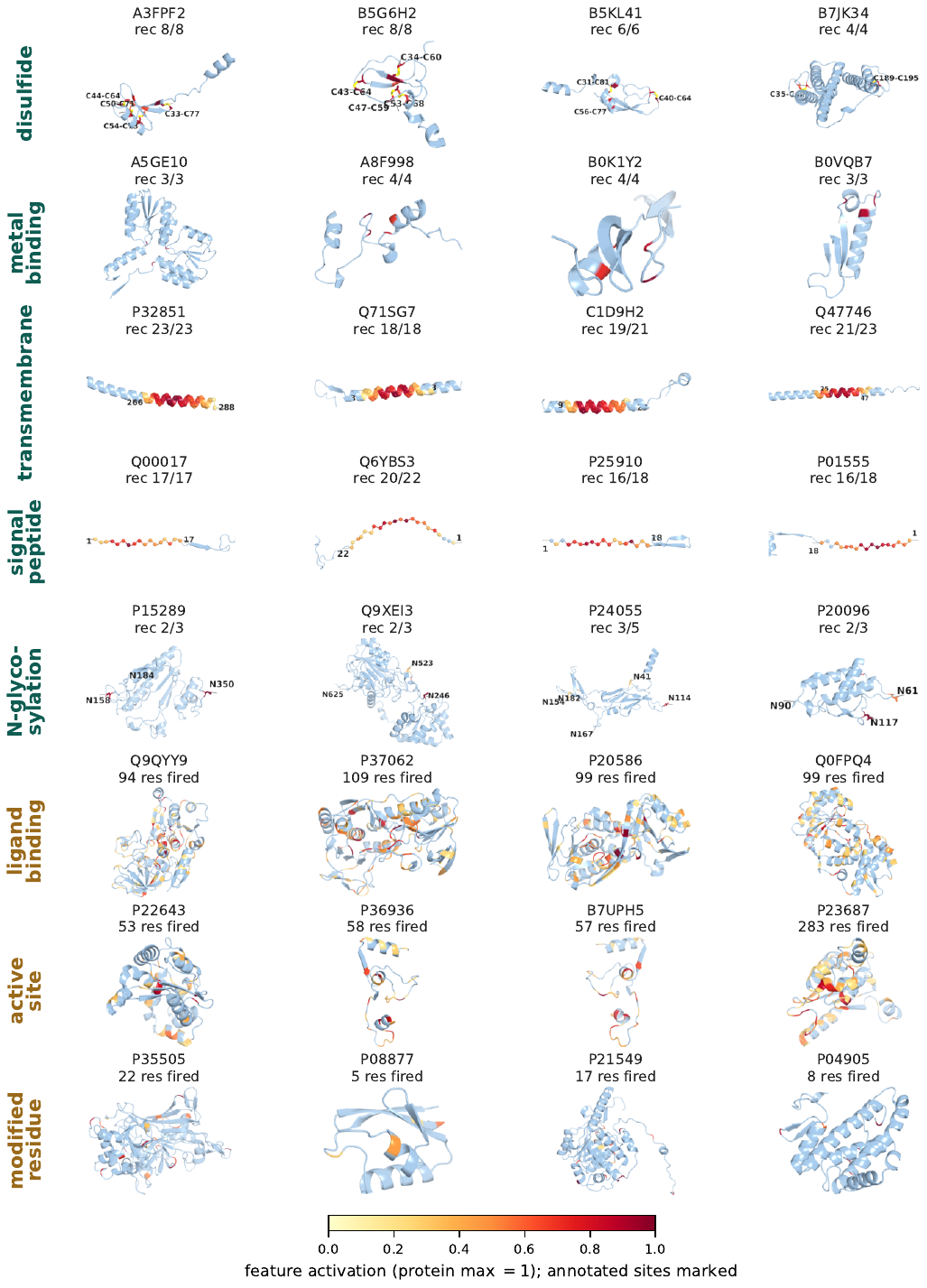}
\caption{Interpreted detectors painted on the folded structure across eight annotation types, four
proteins each. Each panel colors a trunk detector's
per-residue map on the \esmfold{} structure with the annotated sites marked. Green labels mark well-localized features, and yellow labels mark distributed features.}
\label{fig:gradcam_full}
\end{figure*}

\begin{figure*}[t]
\centering
\includegraphics[width=\textwidth]{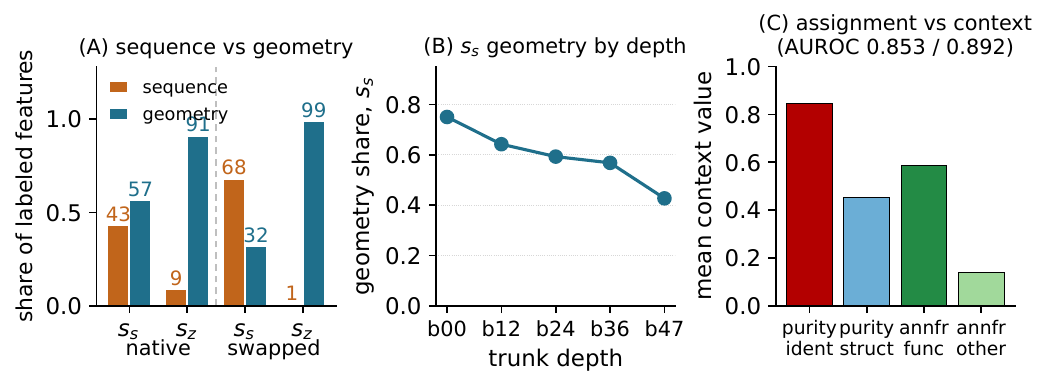}
\caption{The interpreted vocabulary of both representations.
\textbf{(A)}~Share of labeled features that the sequence determines and the share that geometry
determines, for each representation under its own question and under the other's.
\textbf{(B)}~Geometry share of the single-representation vocabulary at each trunk block.
\textbf{(C)}~Amino-acid purity by identity assignment, and top-context annotation density by functional-site assignment.}
\label{fig:autointerp}
\end{figure*}

\begin{figure*}[t]
\centering
\includegraphics[width=\textwidth]{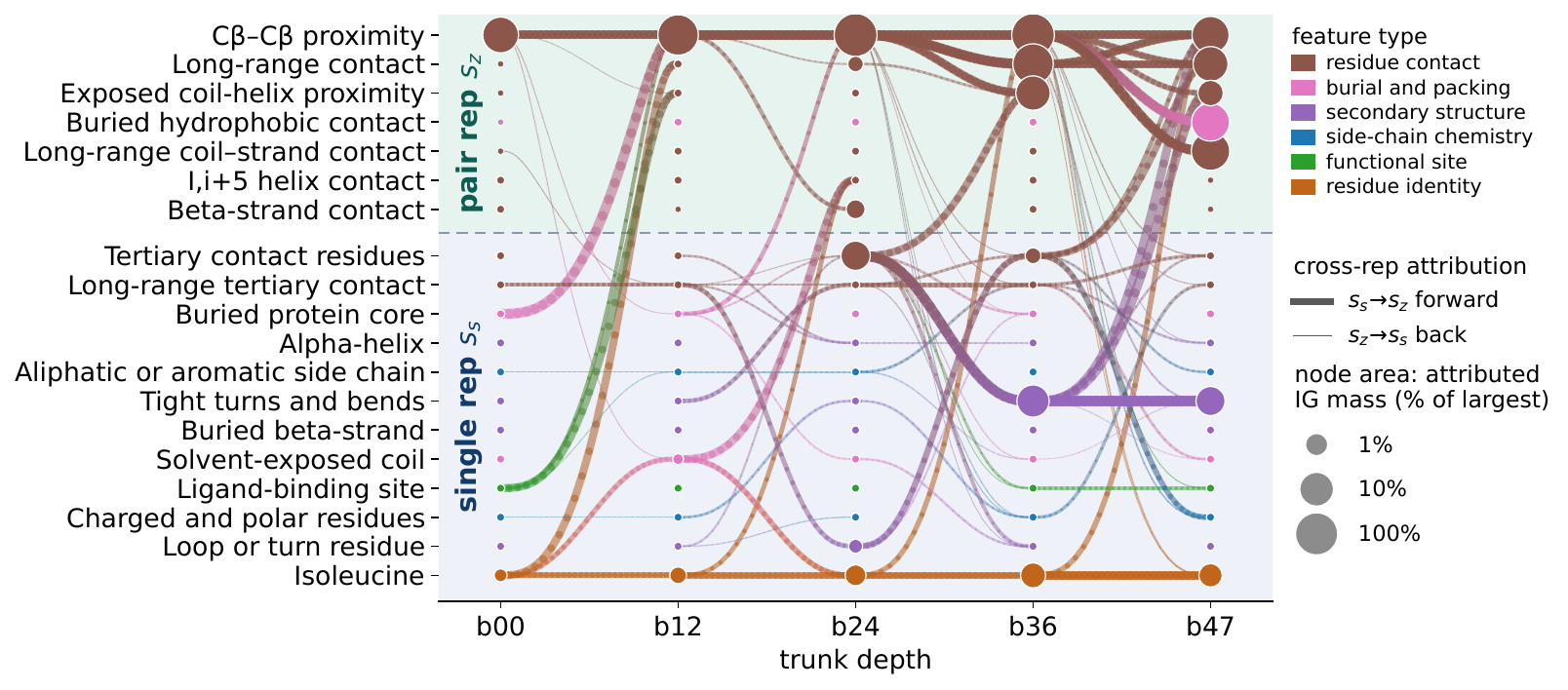}
\caption{The same flow as \Cref{fig:conceptflow} resolved to individual concepts: the twelve largest
single-representation and seven largest pair concepts, with the same two cross-track bundles, forward
$s_s\!\to\!s_z$ thick and backward $s_z\!\to\!s_s$ thin. Node area is attributed mass, edge width
summed weight, both log-scaled.}
\label{fig:conceptflow_full}
\end{figure*}

\Cref{fig:gradcam_full} paints the interpreted detectors on the folded structure across eight
annotation types, four proteins each: the
point-firing disulfide and metal detectors and the span-marking transmembrane, signal, and
N-glycosylation ones localize well, the dense ligand-binding, active-site, and modified-residue
readouts are distributed and less accurate.
\Cref{fig:autointerp} profiles the fixed-taxonomy vocabulary across depth and gives the
label-grounding curves; \Cref{fig:conceptflow_full} extends the category flow of
\Cref{fig:conceptflow} to the individual concepts within the categories. The per-concept circuit is summarized in
\Cref{sec:res-circuits}.

\paragraph{Rendering details of circuit figures.}
\label{app:flowrender}
Rows are annotation categories, and their names in the figure are display forms picked for
legibility. In the single-representation track, the underlying categories are contact, burial,
secondary structure, functional site, and chemical, shown as residue contact, burial and packing,
secondary structure, functional site, and side-chain chemistry. The residue-identity row is the
subset of the chemical category whose cluster terms name one amino acid, separated because it
carries the inherited lexicon rather than a chemical property. The pair-representation rows are the
geometry classes assigned to each pair feature: long-range tertiary, beta-sheet pairing, helix
packing, local/turn, and mixed/other. Every feature enters exactly one row, so the
rows partition the vocabulary.
Every visual channel of the concept-flow figure is a deterministic function of the summed
integrated-gradient weights, disclosed here for reproducibility. Each attributed edge weight is summed
to $(\text{block},\text{upstream concept},\text{downstream concept})$, and a concept node's mass $m$ at
a block is the sum of its incident edge weights. Magnitudes span several orders, so each channel passes
through a log compression to the unit interval,
\begin{equation}
\tilde\phi(w; w_{\max}) = \mathrm{clip}\!\left(
  \frac{\ln w - \ln(0.002\, w_{\max})}{\ln w_{\max} - \ln(0.002\, w_{\max})},\; 0,\; 1\right),
\label{eq:flowscale}
\end{equation}
with $w_{\max}$ the largest value of that channel. Node area is $18 + 430\,\tilde\phi(m; m_{\max})$,
where $m_{\max}$ is the largest node mass across both tracks, so the size legend reads as a percentage of
the largest node. Within-track edges have width $0.5 + 4.5\,\tilde\phi(w; w_{\max})$ and opacity
$0.30 + 0.55\,\tilde\phi(w; w_{\max})$ against the track's largest edge, and their color interpolates
linearly from the source to the destination concept color. The two cross-track bundles are drawn on a
shared scale that encodes their asymmetry: forward $s_s\!\to\!s_z$ edges have width
$0.6 + 4.4\,\tilde\phi(w; w_{\max})$ and backward $s_z\!\to\!s_s$ edges $0.25 + 0.66\,\tilde\phi(w; w_{\max})$,
the backward cap scaled down from the forward cap by the ratio of their shares of the downstream track's
incoming attribution, and the two bow apart laterally at opacities $0.55$ and $0.42$. The figure
aggregates the $28$ largest single-representation and $18$ largest pair-representation concepts by
feature count, and omits within-track edges below $0.002\,w_{\max}$ and
cross-track edges below $0.008\,w_{\max}$.
Areas and widths are read on the log scale of
\eqref{eq:flowscale}.

\section{Controls and robustness}
\label{app:controls}

\paragraph{Confidence intervals and null bands.}
\label{app:ci}
Every confidence interval in this paper is a nonparametric percentile bootstrap: we resample the
evaluation units with replacement under a fixed seed, recompute the statistic on each resample, and
report the $2.5$ and $97.5$ percentiles of the resulting distribution as the $95\%$ interval. Two
resampling units appear. Detection and \withinaa{} selectivity intervals resample the evaluation
proteins as clusters, $1{,}000$ times, so all residues of a protein move together: an overall value
averages the per-protein \auroc{}s of the resampled set, and a \withinaa{} value pools the
per-residue cells of the resampled proteins before averaging, which lets the interval respect the
correlation within a protein. The single-sequence AlphaFold comparison of \Cref{app:aabias}, the
retrieval intervals of \Cref{app:retrieval}, and the output-attribution intervals of
\Cref{sec:res-output} use this same per-protein bootstrap. Causal fold-change intervals instead
resample the attributed targets, $2{,}000$ times, around a ratio of means: each resample divides the
mean intervention effect of the real edges by the mean effect of the matched control, so the reported
multiples, the $1.42\times$ of \Cref{sec:res-circuits} and the $1.64\times$ and $31.7\times$ of
\Cref{app:szcausal}, each carry a percentile interval on that ratio together with the fraction of
targets on which the real edge exceeds its control.

The relational increments of \Cref{tab:relational} are judged against a permutation band in place of
a bootstrap. We permute the trunk feature block across examples under a fixed seed, preserving each
feature's marginal distribution while removing its alignment to residues, then read the \auroc{}
change from appending this misaligned block. The change stays within $\pm 0.010$ on every property.

\paragraph{Feature composition.} The directed edges of \Cref{sec:res-circuits} are established
against matched controls in the native \esmfold{} forward pass. For each attributed edge, we ablate
the upstream feature at its block, on the residues where it is active, and read its attributed
downstream partner at the destination block within the same pass. Over $1{,}063$ edge--protein cases
across the four adjacent block pairs, the ablation moves its partner $1.42\times$ ($95\%$ CI
$[1.38,1.47]$) a write matched on norm and on the signed cosine to the downstream read-out
direction, exceeding that control in $62\%$ of cases. The same intervention moves predicted
geometry, distogram-$L_1$ shifting $3.70\times$ ($[3.37,4.08]$) the matched control. Against a
magnitude-matched control alone, the ratio is $56.6\times$, so magnitude matching is the weaker test
and the composition claim rests on the read-out-matched one. The intervention is small relative to
the representation it displaces, $5.4\%$ of the norm on the residues it touches.
\Cref{app:szcausal} applies this protocol unchanged to the pair representation, where the same
comparison gives $1.64\times$ on $88\%$ of cases, so the pair track carries composition more
strongly than the single one under an identical test.

\paragraph{Output attribution.} At $b_{24}$ the probe of \Cref{sec:res-output} predicts pLDDT and true
per-residue accuracy at $R^2{=}0.73$ and $0.56$, compared with $0.17$ and $0.08$ for a
representation-magnitude baseline at the same block. \esmfold{}'s own confidence and true accuracy agree at
$\rho{=}0.72$. The integrated-gradients attribution is reported as causal mass divided by activation share, so a value above one marks a concept carrying
more causal mass than its share of activation. Ablation of any single feature
produces consistent folding output, so the influence on the
output is distributed across a non-trivial set of features.

\paragraph{Sparsity sweep.} All main-text dictionaries use the sparsest setting, $k{=}16$.
Table~\ref{tab:ksweep} shows reconstruction of the trunk single representation improving
monotonically with the active-feature budget at every block, which we trade for maximally
interpretable codes.

\begin{table}[t]
\centering\small
\setlength{\tabcolsep}{6pt}
\fitwidth{%
\begin{tabular}{lcccc}
\toprule
Block & $k{=}16$ & $k{=}32$ & $k{=}64$ & $k{=}128$ \\
\midrule
b00 & 0.872 & 0.920 & 0.962 & 0.988 \\
b12 & 0.900 & 0.934 & 0.963 & 0.984 \\
b24 & 0.893 & 0.928 & 0.959 & 0.981 \\
b36 & 0.863 & 0.908 & 0.945 & 0.973 \\
b47 & 0.893 & 0.929 & 0.957 & 0.977 \\
\bottomrule
\end{tabular}}
\caption{Reconstruction of the trunk single representation. Reported are explained variance on a held-out set.  We use the sparsest setting, $k{=}16$, trading
reconstruction for maximally interpretable codes.}
\label{tab:ksweep}
\end{table}

\section{Amino-acid type as a measurement bias}
\label{app:aabias}
Amino-acid types encode chemical information and are decodable from the latent, so a detector can
appear structural merely by re-reading residue type. To attribute how much a feature adds
\emph{beyond} amino-acid type, we recompute detection with residue type held fixed. For feature $f$, over
the set $\mathcal{C}$ of (protein, residue-type) cells,
\begin{equation}
\mathrm{AUROC}_{\mathrm{wAA}}(f) = \frac{1}{|\mathcal{C}|}\sum_{(p,r)\in\mathcal{C}}
  \mathrm{AUROC}_{p,r}(z_f),
\label{eq:withinaa}
\end{equation}
where $\mathrm{AUROC}_{p,r}(z_f)$ scores the feature's activation as a detector within residue type
$r$ of protein $p$, so amino-acid type cancels within every cell. Following the control-task
diagnostic~\citep{hewitt2019control}, we read \eqref{eq:withinaa} as attribution across the interpreted
benchmark rather than as a pass/fail test. A feature is scored as a detector over the full sequence of each protein, per protein and then
averaged. We apply a homology filter (5-mer Jaccard $\geq 0.30$) and
protein-stratified splits, and report both micro- and macro-averaged scores with paired
cluster-bootstrap confidence intervals over proteins. Detectors are selected on one split and
scored on the other, and the reverse, so no protein contributes to both selection and evaluation.

Amino-acid type is decodable from the code at \auroc{}$=1.0$, and for several annotations the
residue is the mechanism itself: a disulfide requires cysteine, and N-linked glycosylation requires
the asparagine of its sequon. Holding residue type fixed therefore measures what a feature
contributes beyond that prior: catalytic-site localization holds at \withinaa{} \auroc{}$=0.87$ on
the full Mechanism and Catalytic Site Atlas (M-CSA)~\citep{ribeiro2018mcsa}, and residue burial detection holds at $0.95$.
Amino-acid type acts in two places: on the evaluation of a detector, and on the selection of which feature
becomes the detector. As Table~\ref{tab:aaconfound} reports, choosing by best raw \auroc{} returns a near-pure single-residue
feature whose glycosylation score reads $0.950$ raw and $0.563$ \withinaa{}, while re-selecting
under the control recovers a feature that carries the site beyond amino-acid type ($0.782$) and lifts
active site to $0.974$ and metal binding to $0.924$. Where the annotation is type-independent, the
two rules select the same feature, as for the signal peptide. The gap between them measures how much a
raw-\auroc{} benchmark rewards amino-acid type, and the residual \withinaa{} score measures the structure a
feature carries beyond a legitimate prior.

\begin{table}[t]
\centering\footnotesize
\setlength{\tabcolsep}{1.6pt}
\fitwidth{%
\begin{tabular}{lrrcrrrr}
\toprule
 & & \multicolumn{2}{c}{AA prior} & \multicolumn{2}{c}{raw \auroc{}} & \multicolumn{2}{c}{\withinaa{} \auroc{}} \\
\cmidrule(lr){3-4}\cmidrule(lr){5-6}\cmidrule(lr){7-8}
Property & $n$ & only & res. & naive & ctrl & naive & ctrl \\
\midrule
Disulfide & 771 & 0.995 & C$^{\dagger}$ & 0.956 & 0.892 & 0.768 & \bf 0.908 \\
Glycosyl. & 395 & 0.972 & N$^{\ddagger}$ & 0.950 & 0.691 & 0.563 & \bf 0.782 \\
Active site & 1220 & 0.896 & H & 0.863 & 0.581 & 0.846 & \bf 0.974 \\
Metal bind. & 926 & 0.894 & D & 0.792 & 0.605 & 0.759 & \bf 0.924 \\
Mod. res. & 664 & 0.845 & S & 0.722 & 0.560 & 0.680 & \bf 0.802 \\
Transmem. & 260 & 0.720 & L & 0.744 & 0.572 & 0.777 & 0.771 \\
Signal pep. & 568 & 0.699 & L & 0.881 & 0.881 & 0.965 & 0.965 \\
Ligand bind. & 1525 & 0.673 & G & 0.771 & 0.540 & 0.747 & 0.761 \\
\bottomrule
\end{tabular}}
\caption{Amino-acid type acts both as a biological prior and as a bias on measurement.
Rows are annotated site types. \emph{AA prior} reports the \auroc{} reachable from residue type
alone together with the dominant residue, which the chemistry requires for a disulfide ($\dagger$)
and the sequon for glycosylation ($\ddagger$).}
\label{tab:aaconfound}
\end{table}

\paragraph{Feature-level detection.}
Scored directly on residue annotations, single features localize structural and functional
sites, with disulfide reaching \auroc{}$=0.97$ at the deepest block. A distance-to-centroid
statistic already reaches catalytic sites at \auroc{}$=0.82$ on M-CSA, so a simple geometric prior
recovers much of the raw signal alongside amino-acid type, an effect the protein-SAE literature has left
largely unmeasured. As an external-validity check, the same \withinaa{} catalytic detector applied
to single-sequence AlphaFold structures collapses to \auroc{} between $0.51$ and $0.62$, consistent
with the catalytic code riding the evolutionary signal \esmtwo{} $L_{36}$ supplies rather than
geometry the trunk computes.

\section{Applications of the interpreted detectors}
\label{app:apps}

\begin{table}[t]
\centering\small
\setlength{\tabcolsep}{4pt}
\fitwidth{%
\begin{tabular}{lrrrr}
\toprule
Concept & Feat. & \auroc{} & IoU & \%\,sig. \\
\midrule
Disulfide      & 5934 & 0.98 & 0.79 & 91 \\
Carbohydrate   & 8061 & 0.95 & 0.13 & 52 \\
Active site    & 2819 & 0.80 & 0.03 &  9 \\
Signal peptide & 6323 & 0.79 & 0.35 & 73 \\
Transmembrane  & 5591 & 0.77 & 0.46 & 68 \\
Metal binding  & 1314 & 0.76 & 0.48 & 76 \\
Ligand binding & 2819 & 0.74 & 0.11 & 38 \\
\bottomrule
\end{tabular}}
\caption{Where each interpreted detector reads from. Rows are annotation types; columns give the
feature the detection scan chose, its detection \auroc{}, the mean overlap (IoU) between the
feature's top activating residues and the annotated ones, and the share of proteins where that
overlap beats a residue-permutation baseline.}
\label{tab:gradcam}
\end{table}

Across the seven chosen detectors, the activation sites of \Cref{fig:gradcam} separates point-localized
features from distributed ones, as \Cref{tab:gradcam} reports: disulfide, metal-binding, and transmembrane
features concentrate on their annotated residues, significant against a residue-permutation null on
$68$--$91\%$ of proteins, while the general functional feature shared by the active-site and ligand
annotations spreads across the pocket rather than onto a one- to three-residue point.

\paragraph{Annotating uncharacterized proteins.}
These detectors read the residue-level half of the code, the part inherited from the sequence model,
and they put the interpreted dictionary to work on proteins that carry no labels.
Applied to $300$ human proteins whose UniProt entries carry no functional-site
annotation~\citep{uniprot2023}, the $b_{47}$ detectors, thresholded at a precision operating point
calibrated on annotated proteins, flag sites on $231$ of them. Predicted disulfide cysteines pair spatially,
($594$ of $1{,}121$ flagged cysteines, over $201$ proteins, take a C$\beta$--C$\beta$ partner within
$6.5$\,\AA{} of the predicted fold); predicted signal residues fall in the N-terminal $35$ residues
($634$ of $999$, over $84$ proteins); predicted glycosylation sites land on the N-X-[ST] sequon
($82$ of $154$, over $46$ proteins). In total $127$ proteins carry at least one prediction that
survives its respective sanity check, yielding a concrete, coordinate-level hypothesis about a protein with an empty
annotation record.

\paragraph{Disease enzymes and variant effects.}
On a panel of $139$ brain-disease proteins, with detectors selected out of sample on the
annotation scan, the trunk code localizes annotated sites wherever \esmfold{} is confident
(mean pLDDT $\geq 70$; $236$ of $261$ protein$\times$site-type evaluations): active site
\auroc{}$=0.899$, metal binding $0.815$, ligand binding $0.778$, where a published sequence
dictionary~\citep{simon2024interplm} at $0.650$, $0.543$, $0.574$ and a distance-to-centroid
geometric prior at $0.825$, $0.802$, $0.775$; an evolutionary-conservation baseline stays
competitive throughout ($0.927$, $0.883$, $0.826$). In the low-confidence regions of the panel's
disordered members, among them APP and $\alpha$-synuclein, the trunk score falls to $0.58$--$0.67$,
so the code annotates where a fold exists. Variant effects sit at the edge of what the
code supports: wild-type feature density is at chance on ClinVar
missense variants~\citep{landrum2018clinvar}, while the disruption of the code between wild-type
and mutant reaches \auroc{}${\approx}0.75$, level with the same sequence dictionary scored
identically ($0.731$) and below the masked-marginal sequence-likelihood ceiling
($0.887$)~\citep{meier2021language}. Apparent per-feature $\Delta\Delta G$ prediction dissolves to
\auroc{}$=0.55$ under cross-validation grouped by protein, so the higher plain-split number
reflected protein leakage. The sparse code localizes structure that is folded well,
and variant ranking is behind sequence likelihood.

\section{Circuit attribution}
\label{app:ig}

For a scalar output $F$, an input $x$, and a baseline $x'$, the integrated gradient of input
coordinate $i$ is~\citep{sundararajan2017ig}
\begin{equation}
\mathrm{IG}_i(x) = (x_i - x'_i)\int_0^1
  \frac{\partial F\!\big(x' + \alpha(x - x')\big)}{\partial x_i}\,\mathrm{d}\alpha .
\label{eq:ig}
\end{equation}
For a block pair $(b_{\mathrm{src}},b_{\mathrm{dst}})$ the input is
the upstream code $z$ at $b_{\mathrm{src}}$ and the baseline is the zero code $x'{=}0$, which decodes to
the mean reconstruction. Splicing $\mathrm{decode}(\alpha z)$ into $s_s$ at $b_{\mathrm{src}}$ and
running the forward pass to $b_{\mathrm{dst}}$ yields the downstream code $z^{\mathrm{dst}}(\alpha z)$,
so the code edge \eqref{eq:edge} instantiates \eqref{eq:ig} with the scalar output
$F=z^{\mathrm{dst}}_j$ and baseline $x'{=}0$. We evaluate \eqref{eq:edge} with an $N$-step midpoint
Riemann sum, accumulating
gradient$\times$input online into the per-upstream-feature edge vector at each step so that peak memory
is constant in sequence length, and gradient-checkpoint the spliced span.

\section{Representational similarity between the sequence model and the trunk}
\label{app:cka}
\Cref{fig:prelim}A reports linear CKA over $9{,}000$ residues sampled across $4{,}155$ proteins from
$60$ shards of the corpus; each entry is the mean over five subsamples (maximum standard deviation
$0.012$). CKA is computed in feature space,
$\lVert Y^{\!\top}\! X\rVert_F^2 / (\lVert X^{\!\top}\! X\rVert_F\,\lVert Y^{\!\top}\! Y\rVert_F)$ for
column-centered $X,Y$, agreeing with the Gram form to six decimals. \esmtwo{}--trunk layer pairs average
$0.14$ and \esmtwo{}'s final layer against $b_{00}$ reaches $0.23$, while the trunk blocks agree with
one another at $0.65$ and the middle blocks at $0.96$. Linear CKA is invariant to orthogonal
transforms and isotropic scaling but not to general linear maps, so it captures how the trunk
reorganizes the similarity structure over residues; the decodability probe of \Cref{app:decode} and the
detection results speak to representational content. The block redundancy is what makes five evenly
spaced blocks a sufficient sample of the $48$.

\section{Pairwise distance beyond ESM-2}
\label{app:decode}
For $60{,}000$ residue pairs across $500$ proteins, we regress C$\alpha$--C$\alpha$ distance on each representation. The pair-representation
dictionary code reaches distance $R^2=0.69$ at $b_{47}$ and $0.64$ at $b_{24}$, as \Cref{tab:decode}
reports. No single-residue representation recovers it under the pairwise probe
$[\,s_i;\,s_j;\,s_i\!\odot\! s_j\,]$, including the trunk's own \esmtwo{}-3B backbone. The comparison is scale-matched, and the pair advantage holds at
intermediate blocks that no read-out head reads.
\begin{table}[h]
\centering
\small
\fitwidth{%
\begin{tabular}{lcc}
\toprule
\textbf{predictor} & \multicolumn{2}{c}{\textbf{distance $R^2$}} \\
(representation regressed on distance) & $b_{47}$ & $b_{24}$ \\
\midrule
$s_z$ dictionary code (pair SAE, TopK) & $0.69$ & $0.64$ \\
$s_z$ raw (pair vector, $128$-d) & $0.67$ & $0.62$ \\
\esmtwo{}-3B $L_{36}$, pairwise probe & $-0.11$ & $-0.11$ \\
\esmtwo{}-3B $L_{24}$, pairwise probe & $-0.20$ & $-0.20$ \\
\esmtwo{}-650M $L_{24}$, pairwise probe & $-0.00$ & $-0.00$ \\
sequence separation $|i-j|$ (trivial anchor) & $0.03$ & $0.03$ \\
\bottomrule
\end{tabular}}
\caption{Out-of-sample distance $R^2$ for predicting C$\alpha$--C$\alpha$ distance from each
representation (ridge, protein-stratified split, $60{,}000$ pairs across $500$ proteins). \textbf{predictor}:
the representation supplying the features. \textbf{distance $R^2$}: coefficient of determination on
held-out proteins; higher is better, negative when the fit underperforms predicting the mean. The two
columns index the trunk block supplying the pair code ($b_{47}$, $b_{24}$); the single-residue
baselines do not depend on the trunk block and repeat across columns. \emph{Pairwise probe}
denotes the interaction features $[\,s_i;\,s_j;\,s_i\!\odot\! s_j\,]$ that give a per-residue
representation a pairwise view of the pair $(i,j)$.}
\label{tab:decode}
\end{table}

\section{Recycle similarity}
\label{app:recycle}
\esmfold{} folds by running the trunk for $N{+}1$ passes at $\texttt{num\_recycles}{=}N$ and reading
out the final-recycle state. The single-representation dictionaries are trained on activations
captured at $N{=}1$, while the pair representation, the causal interventions of \Cref{app:szcausal},
and \esmfold{}'s default inference use $N{=}3$. The two settings describe the same representation:
the final-recycle single representation is invariant to the budget, so a dictionary trained under one
reads the other without adjustment.

Refolding the same proteins at $N{=}1$ and $N{=}3$ and capturing the single representation with the
same hook, the two states agree at cosine $1.0000$ at $b_{00}$, $0.99997$ at $b_{24}$, and $0.99979$
at $b_{47}$ (relative $L_2$ $0.15\%$, $0.87\%$, $2.0\%$). Passing through the dictionaries, they select
the same active features on $87$--$98\%$ of residues (TopK Jaccard $0.87$--$0.98$) with activation
correlation $0.98$--$0.99$, the quantity that drives detection. Tracking the representation across
recycles $0$--$4$, it reaches cosine $\geq 0.9998$ of its final value by the first recycle at every
block, and the remaining passes contribute sub-percent refinement. The bulk of the representation is fixed after one recycle and the small
residual carries the functional sharpening, so the detection, within-AA, and circuit analyses read almost identical states across cycles.

% \section{Scientific grounding of the pair circuit}
% \label{app:szscan}
% Each attributed $s_z\!\to\!s_z$ edge is classified against a fixed set of structurally coherent
% transitions: a matched secondary-structure pairing preserved across blocks, local geometry composing
% into longer-range packing, and a disulfide constraint feeding a contact. Of $152$ edges, $120$
% ($79\%$ by count, $88\%$ by attribution weight) fall into these classes; the remainder run the range
% regime backward, a direction recycling permits, and each still connects structurally related concepts. The grounded
% fraction requires a genuine matched secondary-structure pairing rather than any downstream contact.

\section{Causal ablation on pair circuit}
\label{app:szcausal}
The same causal test applies to the pair representation. For each attributed $s_z\!\to\!s_z$ edge we
ablate the upstream pair feature at its block, on the pairs where it is active, and read its
attributed downstream partner at the destination block within the same \esmfold{} forward pass, run
at the model's full recycle budget. Over $1{,}129$ edge--protein cases the ablation moves that
partner $1.64\times$ ($95\%$ CI $[1.58,1.70]$) a write matched on norm and on the signed cosine to
the downstream read-out direction, exceeding that control on $88\%$ of cases; against a
magnitude-matched control alone the ratio is $31.7\times$ ($[29.2,34.3]$), the weaker test as in the
single track. Predicted geometry moves with the intervention, distogram-$L_1$ at the target pairs
shifting $1.81\times$ the matched control.

\section{Protein-level retrieval}
\label{app:retrieval}
We pool each protein's code into a descriptor and retrieve labels by leave-one-out cosine
$k$-nearest-neighbors on top-level enzyme class (EC), fold class (CATH), and superfamily (SCOP),
comparing the trunk code against an identically trained \esmtwo{}-3B dictionary. Pooled per protein,
the two score alike on family labels, and the ordering depends on how the per-residue codes are pooled
into a descriptor.
Across three pooling protocols on the same proteins, concatenated mean and max pooling gives the trunk
$+0.026$ on enzyme class and $-0.008$ to $-0.034$ on the three fold levels; standardizing each feature
before pooling moves every task to between $-0.024$ and $-0.084$; and a binary descriptor that records
which features fire, discarding magnitudes, moves every task to between $+0.047$ and $+0.082$. The
trunk code therefore carries no protein-level advantage that survives the choice of descriptor, and we
draw no claim from it.

Two properties of the benchmark explain why. Family labels propagate by sequence homology, so
retrieval rewards a model trained on evolutionary context: stratifying queries by the sequence
identity of the nearest protein carrying the same label, $2{,}243$ of $2{,}378$ CATH queries have a
relative above $50\%$ identity, and the regime where structural similarity would separate the
two representations is absent from the set. Pooling also discards the pairwise structure that
distinguishes the trunk, collapsing a relation between residues into per-residue marginals, which is
the quantity \Cref{sec:res-relational} measures directly.

\section{Pair-feature label evaluation}
\label{app:labelfirst}
The interpreted labels of the pair features are functionally meaningful at the level of the geometry a
feature carries. We group the pair features by the concept named in their label, blind to any task,
and probe each group on the geometric relations defined from coordinates. Selecting by label
rather than by task performance is the point: a probe trained to predict a relation can succeed on a
leaked correlate, whereas the label was assigned before and independently of any such target. For each
label group we fit a probe on those features alone and read held-out \auroc{} on two relations,
strand pairing and the $\alpha$-helix $i,i{+}4$ contact. Each group predicts its own relation above
the other, as \Cref{tab:labelfirst} reports: $\beta$-labeled features reach $0.938$ on strand pairing
and $0.839$ on the helix contact, while helix-labeled features reverse it, $0.989$ to
$0.737$, with the ordering preserved at $b_{47}$. Both clear a random group of the same size, which
reaches $0.675$ and $0.665$, while the generic-contact group carries the signal the two relations
share and sits between them.

\begin{table}[h]
\centering\small
\setlength{\tabcolsep}{5pt}
\fitwidth{%
\begin{tabular}{lcccc}
\toprule
 & \multicolumn{2}{c}{$b_{24}$} & \multicolumn{2}{c}{$b_{47}$} \\
\cmidrule(lr){2-3}\cmidrule(lr){4-5}
Features labeled & $\beta$-pair & helix $i,i{+}4$ & $\beta$-pair & helix $i,i{+}4$ \\
\midrule
$\beta$-sheet / strand & \bf 0.938 & 0.839 & \bf 0.932 & 0.900 \\
$\alpha$-helix         & 0.737 & \bf 0.989 & 0.834 & \bf 0.993 \\
generic contact        & 0.829 & 0.943 & 0.803 & 0.901 \\
\midrule
random same-size set   & 0.675 & 0.665 & 0.683 & 0.742 \\
full dictionary        & 0.965 & 0.992 & 0.960 & 0.993 \\
\bottomrule
\end{tabular}}
\caption{Held-out \auroc{} for predicting a geometric relation from the pair features carrying a given
label. \textbf{Features labeled}: features grouped by the concept named in their auto-interpretation label,
chosen blind to the target. Columns give the two relations, at two blocks.}
\label{tab:labelfirst}
\end{table}

Each label group predicts its own relation above the other and above a random group of the same size,
a double dissociation that holds at both blocks. The labels therefore track the specific geometry a
feature carries, and the residual cross-prediction is the contact signal the two relations share,
which the generic-contact group carries at both targets.

\section{Prompt inductive bias}
\label{app:crossprompt}
The two representations are labeled with different questions, so the split between them could in
principle follow the question rather than the evidence. We relabel the same features under the other
representation's question, holding the context fixed, and ask whether either question moves a track
toward the other's character.

Our central intuition about the division of labor between single semantic and pair geometric representations holds under swapped prompting. The pair vocabulary is $91\%$ geometric under its own open question and $99\%$
geometric under the single representation's identity-versus-structure question, so asking about
residue identity leaves the pair features reading as geometry. The single vocabulary is $57\%$
geometric under its own question and $32\%$ under the pair representation's relational one, so asking
about spatial relations moves the single features away from geometry rather than toward it. The prompt swapping clears the inductive bias of the prompt on the single and pair representations, verifying the fundamental difference between the information carried by both representations. 

\section{The full attribution}
\label{app:igfull}
\begin{figure*}[p]
\centering
\includegraphics[width=\textwidth]{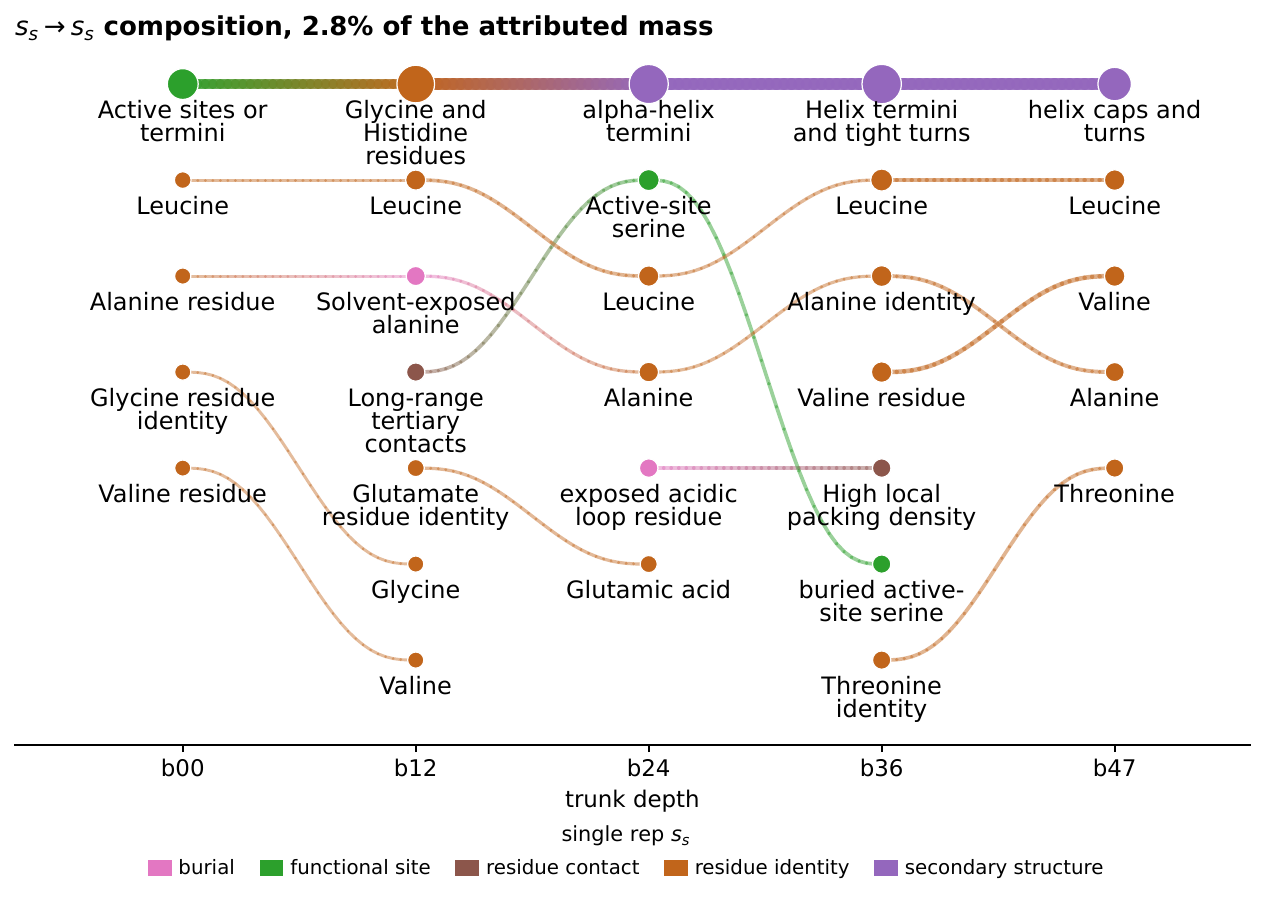}
\caption{The $s_s\!\to\!s_s$ attribution, single-representation composition across depth, at the level of individual features. Nodes are features at a
block, labeled by their own interpretation, sized by attributed mass and colored by category;
edge width is the attributed weight. The title carries this direction's share of the total attributed
mass across the four.}
\label{fig:ig_ss2ss}
\end{figure*}

\begin{figure*}[p]
\centering
\includegraphics[width=\textwidth]{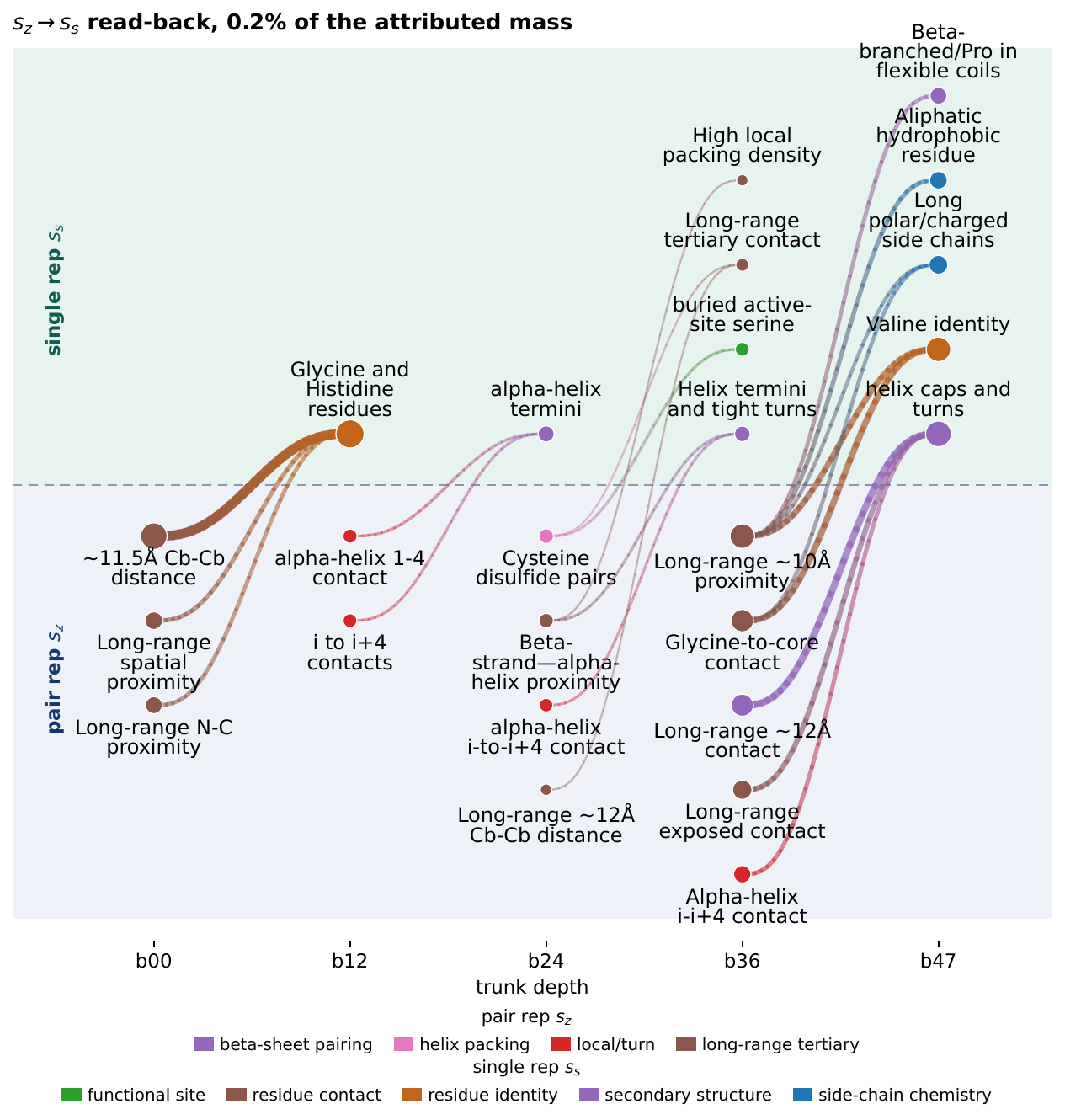}
\caption{The $s_z\!\to\!s_s$ attribution, pair geometry read back into the single code, at the level of individual features. Nodes are features at a
block, labeled by their own interpretation, sized by attributed mass and colored by category;
edge width is the attributed weight. The title carries this direction's share of the total attributed
mass across the four.}
\label{fig:ig_sz2ss}
\end{figure*}

\begin{figure*}[p]
\centering
\includegraphics[width=\textwidth]{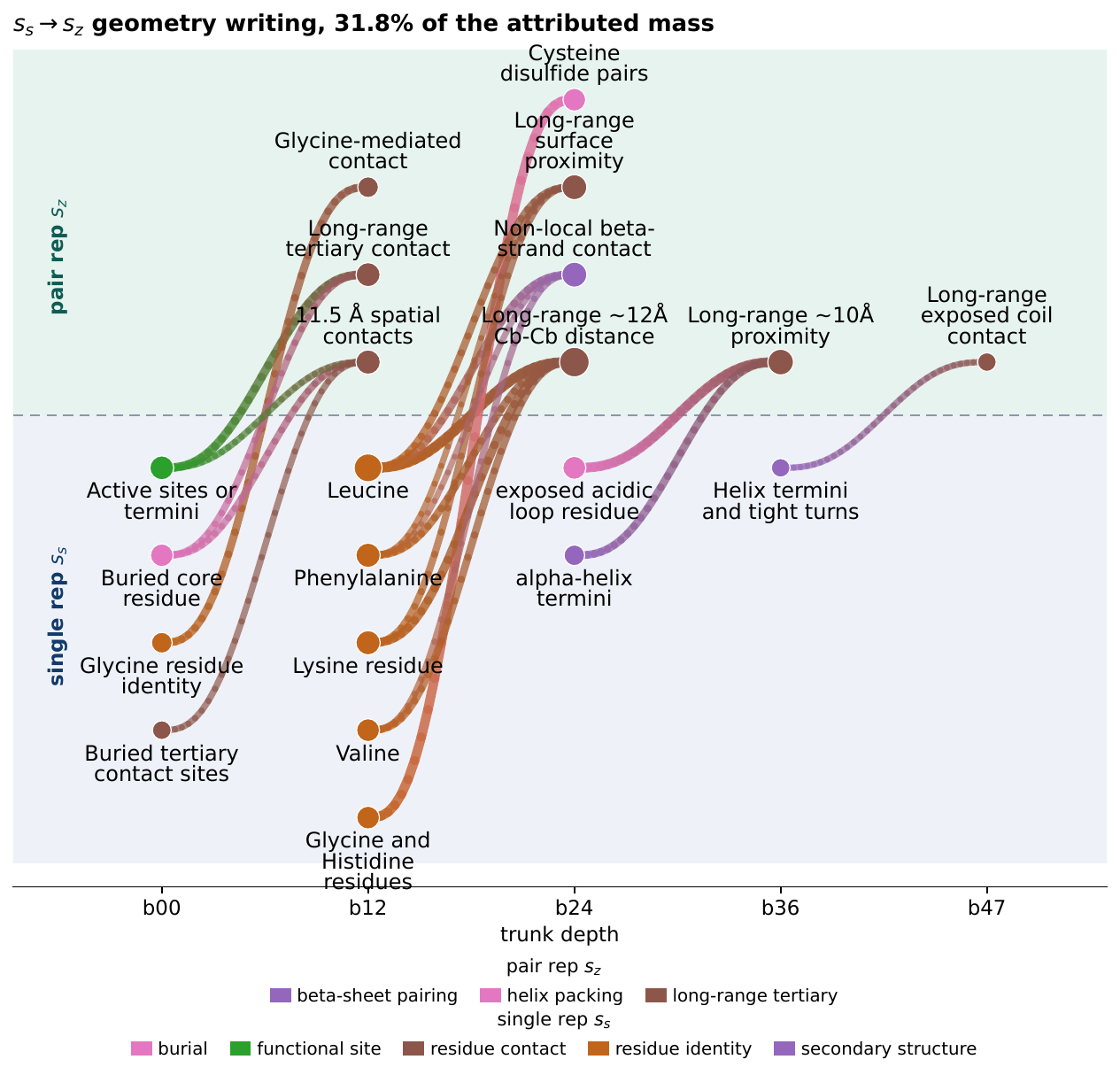}
\caption{The $s_s\!\to\!s_z$ attribution, the single representation writing geometry, at the level of individual features. Nodes are features at a
block, labeled by their own interpretation, sized by attributed mass and colored by category;
edge width is the attributed weight. The title carries this direction's share of the total attributed
mass across the four.}
\label{fig:ig_ss2sz}
\end{figure*}

\begin{figure*}[p]
\centering
\includegraphics[width=\textwidth]{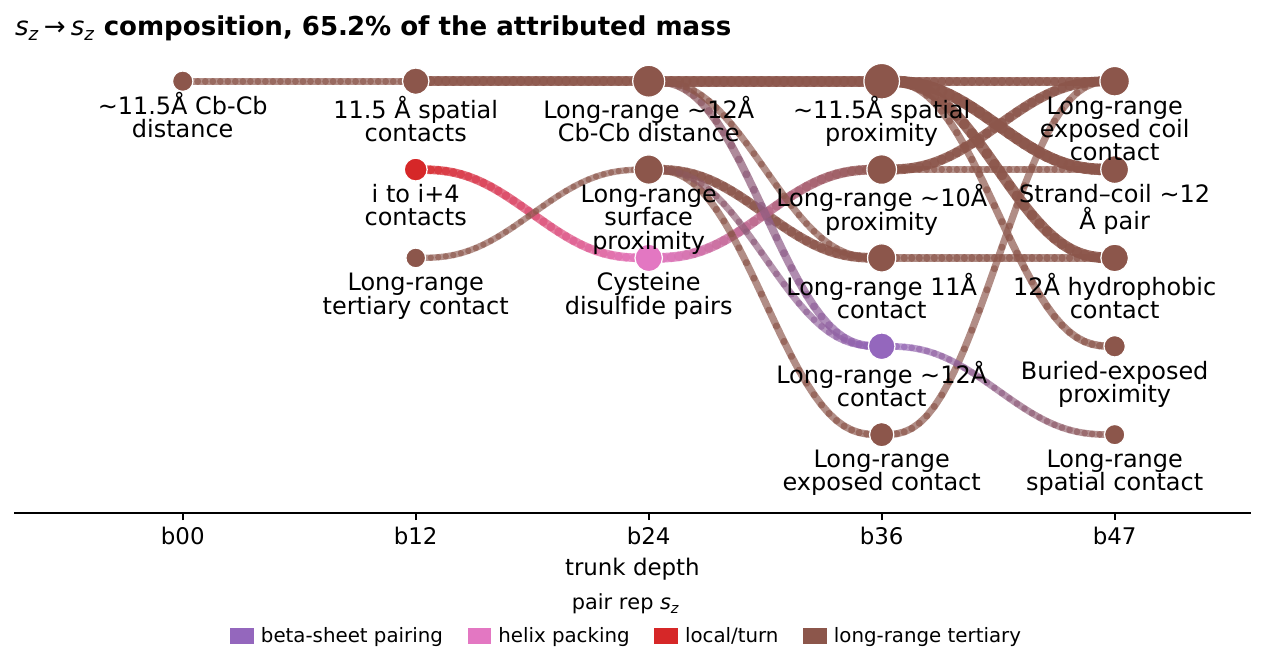}
\caption{The $s_z\!\to\!s_z$ attribution, pair-geometry composition across depth, at the level of individual features. Nodes are features at a
block, labeled by their own interpretation, sized by attributed mass and colored by category;
edge width is the attributed weight. The title carries this direction's share of the total attributed
mass across the four.}
\label{fig:ig_sz2sz}
\end{figure*}

The circuit of \Cref{fig:conceptflow} summarizes four attributions. Each is reported here in full and
at the level of the individual features rather than the categories, so the labels the pipeline wrote
are visible alongside the routes they carry. \Cref{fig:ig_ss2ss,fig:ig_sz2ss} give the two directions
that end in the single representation and \Cref{fig:ig_ss2sz,fig:ig_sz2sz} the two that end in the
pair representation, each across all four block pairs, with the strongest attributed chains drawn. The shares in the titles recover the division of labor the
main text reports: the pair representation accounts for most of the attributed mass, of which the
forward $s_s\!\to\!s_z$ direction supplies about a third, while the backward $s_z\!\to\!s_s$ direction is the
smallest of the four.

\section{Biochemical sanity check}
\label{app:grounding}
Each attributed edge is checked against known biochemistry by asking whether the two interpreted
endpoints have an established relation. The check is a fixed rule set applied automatically to the
label text: an amino-acid-identity endpoint carries a list of
keywords for the geometry its residue is known to favor, a structural endpoint carries the set of
geometry classes its category is known to relate to, and an edge is considered supported when it
matches the opposite endpoint's label. \Cref{tab:rules} gives the cross-track rule set in full.

\begin{table}[h]
\centering\small
\setlength{\tabcolsep}{4pt}
\resizebox{\columnwidth}{!}{
\begin{tabular}{ll}
\toprule
Residue & geometry keywords credited to it \\
\midrule
Gly & turn, local, bend, loop, flex, coil, hinge, termin, cap \\
Pro & turn, local, kink, bend, cap, loop, coil \\
Cys & disulf, cyste, bond, long-range, tertiary \\
Ile, Val & pack, buri, hydroph, sheet, strand, tertiary, long-range, core \\
Leu & pack, buri, hydroph, helix, tertiary, long-range, core \\
Ala & pack, buri, helix, hydroph, core \\
Met & pack, buri, hydroph, core, tertiary \\
Phe, Trp & pack, buri, aromat, stack, hydroph, core, tertiary, long-range \\
Tyr & pack, buri, aromat, stack, hydroph, core, surface, tertiary \\
Asp & turn, local, cap, surface, expos, salt, charg \\
Glu & helix, surface, expos, salt, charg \\
Lys & surface, expos, salt, charg, long-range \\
Arg & surface, expos, salt, charg, long-range, stack \\
His & metal, stack, aromat, coordinat, surface \\
Asn & turn, local, cap, surface, expos, polar \\
Gln & surface, expos, polar, helix \\
Ser & turn, local, cap, surface, polar, hydroxyl \\
Thr & turn, local, cap, strand, sheet, surface, polar \\
\midrule
Category & geometry classes credited to it \\
\midrule
contact & long-range tertiary, beta-sheet pairing, helix packing, local/turn \\
burial & long-range tertiary, helix packing \\
secondary structure & beta-sheet pairing, helix packing, local/turn \\
functional site & long-range tertiary, local/turn \\
disulfide & long-range tertiary \\
\bottomrule
\end{tabular}
}
\caption{The rule set behind the grounding audit. An identity endpoint is credited when the opposite
label contains one of its keywords; a structural endpoint is credited when the pair endpoint's
geometry class appears in its row.}
\label{tab:rules}
\end{table}

An edge between two amino-acid-identity features is grounded
when both read the same residue; an edge joining a residue to a geometry is grounded when that residue
has a documented propensity for it, glycine for turns and local geometry,
cysteine for disulfide pairing, hydrophobic and aromatic residues for packing and buried
contact; an edge between two
structural features is grounded when their categories have an established relation, burial with
packing or secondary structure with strand pairing. A pair endpoint in the catch-all geometry class
carries no specific claim and is excluded rather than counted either way.

The rules support $79\%$ of the edges within either representation, with $609$ of $770$ in the
single track and $120$ of $152$ in the pair one. The grounded pair edges credit genuine matched
secondary-structure relations. 
Across the tracks, the forward $s_s\!\to\!s_z$ leg reaches $75\%$ of edges and $66\%$ of attributed
weight, and the backward $s_z\!\to\!s_s$ leg reaches $63\%$ and $46\%$ of edges and attributed weight, so the leg that carries the least
mass is also the one biochemistry supports least.

The unsupported edges fall into two kinds, and only one of them is plausibly an artifact. Some
carry a local cause into long-range geometry: a helix-termini-and-tight-turns feature writes a
long-range exposed-coil contact, and a loop-and-turn feature writes a long-range contact. Our
criterion counts these as unsupported for a mechanical reason: \Cref{tab:rules} records no relation
between a locally acting category and a fold-spanning geometry class, so any such edge is rejected by
absence. These rejections reflect the limited coverage of the rule set rather than a fault in
the circuit. Others pair a residue with geometry resisted by its chemistry: a helix $i,i{+}4$ contact writes into a valine feature
although $\beta$-branched residues disfavor helix, and a mixed glycine-and-histidine feature writes
into a cysteine disulfide pair, where no
propensity connects the endpoints and the pair label names a residue the upstream feature does not
detect. Throughout, the correspondence we draw is between residue-level causes
and the geometry a trained network computes from them. However, the residue propensities are statistical properties and further verification is needed to accept or reject an attribution edge. 

\section{Auto-interpretation prompt details}
\label{app:prompts}
Each feature is described to the language model by a context block and a question. The two
representations use different context formats, since one activates on residues and the other on
residue pairs, and the questions differ accordingly. Both prompts are reproduced in full in
\Cref{fig:prompt-ss,fig:prompt-sz}, with the evidence block abbreviated to a single exemplar line.
The cross-prompt control of
\Cref{app:crossprompt} exchanges the two questions while holding each context block fixed, so a
single-representation feature is asked the open pair question and a pair feature the
identity-versus-structural question.

\begin{figure*}[!t]
\centering
\begin{minipage}{0.98\textwidth}
\rule{\linewidth}{0.9pt}\\[3pt]
{\footnotesize\textsc{Single-representation prompt}}\\[3pt]
\rule{\linewidth}{0.3pt}\\[2pt]
\small
The feature's top-activating residues, one line per protein:
\begin{small}
\begin{verbatim}
  P02942 pos462 S act=33.4 | SS=alpha-helix ASA=0.06 contacts=6 tert-neigh=10
\end{verbatim}
\end{small}
\vspace{-4pt}
\hspace*{1.2em}$\vdots$\quad\textit{and nine more examples}\\[4pt]
A key defining each field follows, and then the question:\\[3pt]
\texttt{Identify what this feature detects. Distinguish a feature that merely reflects AMINO-ACID
IDENTITY (fires on one residue type regardless of context) from one capturing a STRUCTURAL or
FUNCTIONAL concept (a spatial/functional environment spanning residue types). Ground your label in
the evidence shown; do not infer an annotation that is not present. 2-5 word label.}\\[2pt]
\texttt{JSON: \{"label":"...",\allowbreak{}"explanation":"...",\allowbreak{}"is\_amino\_acid\_identity":bool,\allowbreak{}"is\_structural":bool,\allowbreak{}"category":"secondary\_structure|\allowbreak{}burial|\allowbreak{}contact|\allowbreak{}functional\_site|\allowbreak{}disulfide|\allowbreak{}chemical|\allowbreak{}positional|\allowbreak{}other"\}}\\[3pt]
\rule{\linewidth}{0.9pt}
\end{minipage}
\caption{The single-representation interpretation prompt. Line breaks within the question are set by
the page, not by the prompt, which supplies the text as one paragraph followed by the schema line.}
\label{fig:prompt-ss}
\end{figure*}

\begin{figure*}[!t]
\centering
\begin{minipage}{0.98\textwidth}
\rule{\linewidth}{0.9pt}\\[3pt]
{\footnotesize\textsc{Pair-representation prompt}}\\[3pt]
\rule{\linewidth}{0.3pt}\\[2pt]
\small
Summary statistics over the feature's activating pairs open the block: the count of pairs and
proteins, median C$\beta$--C$\beta$ distance and sequence separation, the long-range and local
fractions, the fraction with both endpoints buried, the fraction carrying an annotation, and the
dominant secondary-structure pair. Exemplar pairs follow:
\begin{small}
\begin{verbatim}
  pair 69-120 (L-W) dist=4.8A sep=51 | SS=E/E ASA=0.00/0.26 tert=14/12 | i:- j:- | act=42.4
\end{verbatim}
\end{small}
\vspace{-4pt}
\hspace*{1.2em}$\vdots$\quad\textit{and four other}\\[4pt]
The question leaves the answer open rather than presupposing geometry:\\[3pt]
\texttt{What does this feature encode? Judge the distribution; use exemplars as evidence. Weigh: a
spatial relationship between i and j; the amino-acid types involved regardless of geometry; a
sequence-offset or position pattern; a functional pairing; a property of each endpoint alone that
says nothing about the relationship; or no consistent pattern. Do not assume it is geometric. Answer
only with the JSON, briefly.}\\[2pt]
\texttt{JSON: \{"label":"2-5 words",\allowbreak{}"explanation":"one sentence, name the runner-up you rejected",\allowbreak{}"is\_amino\_acid\_identity":bool,\allowbreak{}"is\_structural":bool,\allowbreak{}"category":"...",\allowbreak{}"pair\_relation":"short phrase",\allowbreak{}"range\_regime":"local|\allowbreak{}medium|\allowbreak{}long"\}}\\[3pt]
\rule{\linewidth}{0.9pt}
\end{minipage}
\caption{The pair-representation interpretation prompt. As above, the question is one paragraph in
the prompt; the breaks shown are the page's.}
\label{fig:prompt-sz}
\end{figure*}

\section{Splitting the vocabulary into sequence and geometry}
\label{app:seqgeom}
A category counts as sequence-determined when it names residue identity or a sequence offset, which
covers the side-chain-chemistry and sequence-position categories, and as geometric when it names a
relation between residues or a 3D environment, which covers residue contact and
burial. Three categories fall cleanly on neither side: secondary structure, which a sequence model already
predicts well although it describes backbone geometry; functional site, which is annotated per
residue although the site is a spatial arrangement; and disulfide, which requires two cysteines in
sequence and a bond in space. Those are excluded from both sides rather than assigned, $2{,}097$ of
the $5{,}595$ single-representation features and $350$ of the $2{,}972$ pair features, so the shares
reported are over the features whose category settles the question. The counts behind
\Cref{fig:autointerp}A are $3{,}498$ and $2{,}622$ features under the native questions and $662$ and
$906$ under the swapped ones.

\end{document}